\documentclass[twocolumn]{article}

\usepackage{amssymb,amsmath,latexsym,amsthm,amsfonts}
\usepackage{graphicx}
\usepackage{mathrsfs}
\usepackage{natbib}
\bibpunct{(}{)}{,}{a}{}{,}

\usepackage{color}

\newcommand\ds{\displaystyle}
\newcommand\Tee{\mathrm{T}}

\newcommand\bx{\mathbf{x}}
\newcommand\bX{\mathbf{X}}
\newcommand\by{\mathbf{y}}
\newcommand\bz{\mathbf{z}}
\newcommand\btheta{\boldsymbol{\theta}}
\newcommand\bbeta{\boldsymbol{\beta}}
\renewcommand\P{\mathbb{P}}

\newtheorem{example}{Example}{\bfseries}{\sffamily}
\newcommand\findeX{\hfill$\blacktriangleleft$\end{rm}}

\title{{\bf On computational tools for Bayesian data analysis}}

\author{{\sc Christian P.~Robert and Jean-Michel Marin}\thanks{C.P. Robert is Professor of Statistics at
Universit\'e Paris-Dauphine, CEREMADE, 75775 Paris cedex 16, and Head of the Statistics Lab at
CREST, INSEE, France.  Email: \texttt{xian@ceremade.dauphine.fr} Webpage:
\texttt{www.ceremade.dauphine.fr/$\sim$xian} Blog: \texttt{xianblog.wordpress.com} 
Jean-Michel Marin is Professor of Statistics at Institut de Math\'ematiques et Mod\'elisation de Montpellier,
Universit\'e Montpellier 2, Case Courrier 51 34095 Montpellier cedex 5, and associate researcher at CREST, INSEE, Paris,
Email: \texttt{jean-michel.marin@math.univ-montp2.fr}}
}

\begin{document}

\twocolumn
\maketitle

\begin{abstract}
While the previous chapter (Robert and Rousseau, 2010) addressed the foundational aspects of Bayesian analysis, the
current chapter details its practical aspects through a review of the computational methods
available for approximating Bayesian procedures. Recent innovations like Monte Carlo
Markov chain, sequential Monte Carlo methods and more recently Approximate Bayesian Computation techniques
have considerably increased the potential for Bayesian applications and they have consequently opened new avenues
for Bayesian inference, first and foremost Bayesian model choice. 

\vspace{0.25cm} \noindent
\end{abstract}

\section{Introduction}

The previous chapter (Robert and Rousseau, 2010) has (hopefully) stressed the unique coherence
of Bayesian data analysis ---the complete inferential spectrum (estimators, predictors, tests, confidence 
regions, etc.) is derived from a unique perspective, once both a loss function and a prior distribution are 
constructed---, but it has not adressed the complex issues related to the practical implementation  of this
analysis that usually involves solving integration, optimisation and implicit equation
problems, most often simultaneously.  

This computational challenge offered by Bayesian inference has led to a specific branch of Bayesian statistics
concerned with these issues, from the early approximations of Laplace to the numerical probability developments
of the current days. In particular, the past twenty years have witnessed a tremendous surge in computational
Bayesian statistics, due to the introduction of powerful approximation methods like Markov chain
(MCMC) and sequential Monte Carlo techniques. To some extent, this branch of Bayesian statistics
is now so intricately connected with Bayesian inference that some notions like Bayesian model choice and Bayesian model 
comparison hardly make sense without it. 

The probabilistic nature of the objects, involved in those computational challenges, as well as their 
potentialy high dimension, led the community to opt for simulation based, rather than numerical, solutions.
While numerical techniques are indeed used to solve some optimisation or some approximation setups, even
producing specific approaches like variational Bayes \citep{jaakkola:jordan:2000}, the method of choice is
simulation, i.e.~essentially the use of computer generated random variables and the reliance on the Law of
Large Numbers. For instance, all major softwares that have been built towards Bayesian data analysis like {\sf WinBUGS} 
and {\sf JAGS}, are entirely depending upon simulation approximations. We will therefore abstain from describing any 
further the numerical advances found in this area, 
referring the reader to \cite{spall:2003} and \cite{gentle:2009} for proper coverage. 

In this chapter, we thus discuss simulated-based computational methods in connection with a few model choice examples 
(Section \ref{sec:chill}), separating non-Markovian (Section \ref{sec:moca}) from Markovian 
(Section \ref{sec:mcmc}) solutions. For detailed entries on Bayesian computational statistics, 
we refer the reader to \cite{Chen:Shao:Ibrahim:2000}, \cite{Liu:2001} or 
\cite{Robert:Casella:2004,robert:casella:2009}, pointing out that 
books like \cite{albert:2009} and \cite{marin:robert:2007} encompass both Bayesian inference
and computational methodologies in a single unifying perspective.

\section{Computational difficulties}\label{sec:chill}

In this section, we consider two particular types of statistical models with computational
challenges that can only be processed via simulation.

\subsection{Generalised linear models}

Generalised linear models \citep{McCullagh:Nelder:1989} are extensions of the standard linear regression model.
In particular, they bypass the compulsory selection of a single transformation of the data that must achieve
the possibly conflicting goals of normality and linearity, goals which are imposed by the linear regression model
but that are impossible to achieve for binary or count responses.

Generalised linear models formalise the connection between a response variable $y\in\mathbb{R}$ and a vector $\bx\in\mathbb{R}^p$
of explanatory variables. They assume that the dependence of $y$ on $\bx$ is partly linear in the sense that the conditional
distribution of $y$ given $\bx$ is defined in terms of a linear combination $\bx^\Tee\bbeta$ of the
components of $\bx$ ($\bx^\text{T}$ being the transpose of $\bx$),
$$
y|\bx,\bbeta \sim f(y|\bx^\Tee\bbeta)\,.
$$

We use the notation
$\by=\left(y_1,\ldots,y_n\right)$ for a sample of $n$ responses and
 $$
\bX=\left[\bx_1\quad\ldots\quad \bx_p\right]=
$$
$$
\left[\begin{array}{cccc}
 x_{11} & x_{12} & \ldots & x_{1p} \\
 x_{21} & x_{22} & \ldots & x_{2p} \\
 x_{31} & x_{32} & \ldots & x_{3p} \\
 \vdots & \vdots & \vdots & \vdots \\
 x_{n1} & x_{n2} & \ldots & x_{np}
\end{array}\right]=\left[\begin{array}{c}
 \bx^1 \\
 \vdots \\
 \bx^n
\end{array}\right]
$$
for the $n\times p$ matrix of corresponding explanatory variables,
possibly with $x_{11}=\ldots=x_{n1}=1$ ($y$ and $\bx$ correspond to
generic notations for single-response and covariate vectors,
respectively).

A generalized linear model is specified by two functions:
\begin{itemize}
\item[(i)] a conditional density $f$ on $y$ conditional on $\bx$ that belongs to an exponential family 
and that is parameterized by an expectation parameter
$\mu=\mu(\bx)=\mathbb{E}[y|\bx]$ and possibly a dispersion parameter $\phi>0$ that does 
not depend on $\bx$; and
\item[(ii)] a link function $k$ that relates the mean $\mu=\mu(\bx)$ of $f$
and the covariate vector, $\bx$, through $k(\mu)=(\bx^\Tee\bbeta)$, $\bbeta\in\mathbb{R}^p$.
\end{itemize}
For identifiability reasons, the link function $k$ is a one-to-one function and we have
$$
\mathbb{E}[y|\bx,\bbeta,\phi]=k^{-1}\left(\bx^\Tee\bbeta\right)\,.
$$
We can thus write the (conditional) likelihood as
$$
\ell(\bbeta,\phi|\by,\bX)=\prod_{i=1}^n f\left(y_i|\bx^{i\Tee}\bbeta,\phi\right)\,.
$$
In practical applications like econometrics or genomics, $p$ can be very large and even larger 
than the number of observations $n$. Bayesian data analysis on $\bbeta$ and possibly $\phi$
proceeds through the posterior distribution of $(\bbeta,\phi)$ given $(\bX,\by)$:
\begin{equation}
\pi(\bbeta,\phi|\bX,\by) \propto \prod_{i=1}^n f\left(y_i|\bx^{i\Tee}\bbeta,\phi\right) \,\pi(\bbeta,\phi|\bX) \label{eq:glm}
\end{equation}
which is never available as a standard distribution outside the normal linear model. 
Indeed, the choice of the prior distribution $\pi(\bbeta,\phi|\bX)$ depends on the prior information available 
to the modeller. In cases when $\phi=1$, we will use the default solution advocated in \cite{marin:robert:2007}, namely the extension
of Zellner's $g$-prior that was originaly introduced for the linear model, as discussed in the previous chapter:
$$
\bbeta|\bX \sim \mathcal{N} \left( 0, n \left(\bX^\Tee\bX\right)^{-1}\right)\,.
$$
The motivation behind the factor $n$ is that the information brought by the prior is scaled to
the level of a single observation. Even this simple modeling does not avoid the computational issue
of exploiting the posterior density \eqref{eq:glm}.

\begin{example}\label{ex:glum}\begin{rm} 
A specific if standard case of generalised linear model for binary data is the {\em probit model}:
$Y(\Omega)=\{0,1\}$ and we have
$$
\P(Y=1|\bx) = 1 - \P(Y=0|\bx) = \Phi(\bx^\Tee\beta)\,,
$$
where $\Phi$ denotes the standard normal cumulative distribution function. Under the $g$-prior
$\pi(\bbeta|\bX)$ presented above, the corresponding posterior distribution, proportional to
\begin{equation}\label{eq:propos}
\pi(\bbeta|\bX)\,\prod_{i=1}^n \Phi(\bx^{i\Tee}\bbeta)^{y_i} \Phi(-\bx^{i\Tee}\bbeta)^{1-y_i}\,,
\end{equation}
is available in closed form, up to the normalising constant, but
is not a standard distribution and thus cannot be easily handled!

In this chapter, we will use as illustrative data the Pima Indian diabetes study available in
R \citep{rmanual} as the {\sf Pima.tr} dataset with 332 women registered
and consider a probit model predicting the presence of diabetes from three predictors,
the glucose concentration (glu), the diastolic blood (bp) pressure and the diabetes pedigree function (ped),
$$
\P(y=1|\bx) = \Phi(x_1\beta_1+x_2\beta_2+x_3\beta_3)\,.
$$
A maximum likelihood estimate of the regression coefficients is provided by {\sf R glm} function
as\index{glm@\verb+glm+}

\footnotesize \begin{verbatim}
Deviance Residuals: 
    Min       1Q   Median       3Q      Max  
-2.1347  -0.9217  -0.6963   0.9959   2.3235  
Coefficients:
     Estimate Std. Error z value Pr(>|z|)    
glu  0.012616   0.002406   5.244 1.57e-07 ***
bp  -0.029050   0.004094  -7.096 1.28e-12 ***
ped  0.350301   0.208806   1.678   0.0934 .  
---
Signif. codes: '***' 0.001 '**' 0.01 '*' 0.05 '.' 0.1  

Null deviance: 460.25  on 332  degrees of freedom
Residual deviance: 386.73  on 329  degrees of freedom
AIC: 392.73
Number of Fisher Scoring iterations: 4
\end{verbatim} 
\normalsize the final column of stars indicating a possible significance of the first two covariates from a classical
viewpoint.
\findeX\end{example}

This type of model is characteristic of conditional models where there exist a plethora of covariates 
$x_i$---again, potentially more than there are observations---and one of the strengths of Bayesian 
data analysis is to be able to assess the impact of those covariates on the dependent variable $y$. 
This obviously is a special case of model choice, where a given set of covariates is associated
with a specific model. As discussed in the previous chapter, the standard Bayesian solution in
this setting is to compute posterior probabilities or Bayes factors for all models in competition.
For instance, if a single covariate, $x_3$ (ped) say, is under scrutiny, the Bayes factor associated with
the null hypothesis $H_0:~\beta_3=0$ is
\begin{equation}\label{Baf}
B^\pi_{01}=\frac{m_0(\by)}{m_1(\by)}
\end{equation}
where $m_0$ and $m_1$ are the marginal densities under the null and the alternative 
hypotheses,\footnote{As will become clearer in Section \ref{Bprox}, Bayes factor approximations
are intrinsically linked with the normalising constants of the posterior distributions of
the models under competition. We already stressed in Robert and Rousseau (2010) that this is a 
special case when normalising constants matter!} i.e.
$$
m_i(\by) = \int f(\by|\bbeta,\bX_i) \pi_i(\bbeta|\bX_i) \text{d}\bbeta\,,
$$
$\pi_0$ being the $g$-prior excluding the covariate $x_3$. 

If we denote by
$\bX_0$ the $332\times 2$ matrix containing the values of \verb+glu+ and \verb+bp+
for the $332$ individuals and by $\bX_1$ the $332\times 3$ matrix containing the values of
the covariates \verb+glu+, \verb+bp+ and \verb+ped+, the Bayes factor $B_{01}$ is given by

{\scriptsize{ 
\begin{align*}
&(2\pi)^{1/2}n^{1/2}\frac{|(\bX_0^\text{T}\bX_0)|^{-1/2}}{|(\bX_1^\text{T}\bX_1)|^{-1/2}}\times\\
&\frac{\ds \int_{\mathbb{R}^2}\prod_{i=1}^n \frac{\{1-\Phi\left((\bX_0)_{i,\cdot}\bbeta\right)\}^{1-y_i}}
{\Phi\left((\bX_0)_{i,\cdot}\bbeta\right)^{-y_i}}
\exp\left\{-\bbeta^\text{T}(\bX_0^\text{T}\bX_0)\bbeta/2n\right\}\text{d}\bbeta}
{\ds \int_{\mathbb{R}^3}\prod_{i=1}^n \frac{\{1-\Phi\left(\bX_1)_{i,\cdot}\bbeta\right)\}^{1-y_i}}
{\Phi\left(\bX_1)_{i,\cdot}\bbeta\right)^{-y_i}}
\exp\left\{-\btheta^\text{T}(\bX_1^\text{T}\bX_1)\bbeta/2n\right\}\text{d}\bbeta}
\end{align*}}}\normalsize
using the shortcut notation that $A_{i,\cdot}$ is the $i$-th line of the matrix $A$.

The approximation of those marginal densities,
which are not available outside the normal model \citep[see, e.g.,][Chapter 3]{marin:robert:2007},
is thus paramount to decide about the inclusion of available covariates.

In this setting of selecting covariates in a conditional model, an additional and non-negligible
computational difficulty is that the number of hypotheses to be tested is $2^p$ if each of
the $p$ covariates is under scrutiny. When $p$ is large, it is simply impossible to envision all possible
subsets of covariates and a further level of approximation must be accepted, namely that only the
most likely subsets will be visited by an approximation method. 

\subsection{Challenging likelihoods}\label{sec:watsup}
A further degree of difficulty in the computational processing of Bayesian models
is reached when the likelihood function itself cannot be computed in a
reasonable amount of time. Examples abound in econometrics, physics, astronomy, genetics, and beyond.
The level of difficulty may be that the computation time of one single value of the likelihood function 
requires several seconds, as in the cosmology analysis of \cite{wraith-2009-80} where the likelihood
is represented by an involved computer program. It may also be that the only possible representation
of the likelihood function is as an integral over a possibly large number of latent variables of a
joint (unobserved) likelihood. 

\begin{example}\label{probitatent}\begin{rm} (Continuation of Example \ref{ex:glum})
Although the likelihood of a probit model is available in closed form, this
probit model can be represented as a natural latent variable model. If
we introduce an artificial sample $\bz=(z_1,\ldots,z_n)$ of $n$ independent latent variables
associated with a standard regression model, i.e.~such that
$z_i|\bbeta\sim\mathcal{N}\left(\bx^{i\Tee}\bbeta,1\right)$,
where the $\bx^{i\Tee}$'s are the $p$-dimensional covariates and $\bbeta$ is the vector of
regression coefficients, then $\by=(y_1,\ldots,y_n)$ defined by
$
y_i = \mathbb{I}_{z_i>0}
$
is a probit sample. Indeed, given $\bbeta$, the $y_i$'s are independent Bernoulli rv's with
$\P(Y_i=1|\bx^i,\bbeta)=\Phi\left(\bx^{i\Tee}\bbeta\right)$.
\findeX\end{example}

Such latent variables models are quite popular in most applied fields.
For instance, a stochastic volatility model \citep{jacquier:polson:rossi:1994,chib:02} includes
as many (volatility) latent variables as observations. In a time series with thousands of periods, this
feature means a considerable increase in the complexity of the problem, as the volatilities cannot be 
integrated analytically and thus need to be simulated. Similarly, phylogenetic trees (REF) that reconstruct
ancestral histories in population genetics are random trees and a nuisance parameter for inference about
evolutionary mechanisms, but, once more, they cannot be integrated.

\begin{example}\label{openup}\begin{rm} Capture-recapture experiments are used in ecology to 
assess the size and the patterns of a population of animals by a series of captures where
captured animals are marked, i.e.~individualy identified as having been captured once, and
released. The occurence of recaptures is then informative about the whole population. A longer
description is provided in Marin and Robert (2007, Chapter 5), but we only consider here a three stage
{\em open population} capture-recapture model, where there is a probability $q$ for each individual in the population
to leave the population between each capture episode. Due to this possible emigration of animals,
the associated likelihood involves unobserved indicators and we study here the case where only the 
individuals captured during the first capture experiment are marked and subsequent
recaptures are registered. This model is thus described via the summary statistics
\begin{align*}
&n_1\sim \mathscr{B}(N,p)\,,\quad
r_1|n_1\sim\mathscr{B}(n_1,q)\,,\\
&r_2|n_1,r_1\sim\mathscr{B}(n_1-r_1,q)\,,\quad
c_2|n_1,r_1\sim\mathscr{B}(n_1-r_1,p),\,
\end{align*}
and
$$
c_3|n_1,r_1,r_2\sim\mathscr{B}(n_1-r_1-r_2,p)\,,
$$
where only the first capture size, $n_1$, the first recapture size, $c_2$, and 
the second recapture size, $c_3$, are observed. The numbers of marked individuals removed at stages
$1$ and $2$, $r_1$ and $r_2$, are not observed and are therefore latent variables of the model.
If we incorporate those missing variables within the parameters,
the likelihood $\ell(N,p,q,r_1,r_2|n_1,c_2,c_3)$ is given by
\begin{align*}
{N\choose n_1}&p^{n_1}(1-p)^{N-n_1}\,{n_1\choose r_1}\,
        q^{r_1}(1-q)^{n_1-r_1}\\
&\times{n_1-r_1\choose c_2}\,p^{c_2}(1-p)^{n_1-r_1-c_2}\\
&\times{n_1-r_1\choose r_2} q^{r_2}(1-q)^{n_1-r_1-r_2}\\
&\times{n_1-r_1-r_2\choose c_3} p^{c_3}(1-p)^{n_1-r_1-r_2-c_3}
\end{align*}
and, if we use the improper prior $\pi(N,p,q)=N^{-1}\mathbb{I}_{[0,1]}(p)\mathbb{I}_{[0,1]}(q)$, the posterior
on the $(N,p,q,r_1,r_2|n_1,c_2,c_3)$ is available up to a constant. Summing over all possible values of
$(r_1,r_2)$ to obtain the posterior associated with the ``observed" likelihood creates some difficulties
when $n_1$ is large. Indeed, this summation typically introduces a lot of numerical errors.

The dataset associated with this example is extracted from Marin and Robert's 
(\citeyear{marin:robert:2007}, Chapter 5) {\em eurodip} dataset and is related 
to a population of birds called {\em European dippers}.\footnote{European dippers are 
strongly dependent on streams, feeding on underwater invertebrates, and
their nests are always close to water. The capture--recapture data 
contained in the {\em eurodip} dataset covers 7 years of observations
in a zone of $200$ $\text{km}^2$ in eastern France.} For the 1981 captures,
we have $n_1=22$, $c_2=11$, and $c_3=6$.
\findeX\end{example}

The following example is a different case where the likelihood is missing a term that cannot be reconstituted by
completion and thus requires a custom-built solution.

\begin{example}\label{ex:sherlknn}\begin{rm} The {\em $k$-nearest-neighbour} procedure is a classification
procedure that uses a training dataset $(y_i,\mathbf{x}_i)_{1\le i\le n}$ for prediction purposes.
The observables $y_i$ are class labels, $y_i\in \{1,\ldots,G\}$, while the $\mathbf{x}_i$ are
covariates, possible of large dimension. When observing a new covariate $\mathbf{x}_{n+1}$, the
corresponding unobserved label $y_{n+1}$ is predicted as the most common class label found in 
the $k$ nearest neighbours of $\mathbf{x}_{n+1}$ in $\bX=\{\mathbf{x}_{1},\ldots,\mathbf{x}_{n}\}$,
the neighbours of a covariate vector being defined by the usual Euclidean norm. 
\cite{Cucala:Marin:Robert:Titterington:2006} have proposed a probabilistic model for this classification 
mechanism. They first propose to model the distribution of $\by$:
\begin{align}
&f(\by|\bX,\beta,k)=\nonumber\\
&{\ds \exp\left(\beta \sum_{i=1}^n \sum_{\ell\,\sim_k i} 
\delta_{y_i}(y_\ell)\bigg/ k\right)}\Bigg/{\ds Z(\beta,k)}
\label{eq:knn}
\end{align}
where $\delta_x(y)$ denotes the Kroenecker delta, $Z(\beta,k)$ is the normalising constant of the density
and where ${\ell\sim_k i}$ means that the summation is taken over the observations $\bx_i$
for which $\bx_\ell$ is a $k$-nearest neighbour. The motivation for this
modelling is that the full conditionals corresponding to \eqref{eq:knn} are given by
\begin{align}
&f(y_i|\by_{-i},\bX,\beta,k)\propto\nonumber\\
&\exp\left\{ \beta/k \left(\sum_{\ell\,\sim_k i} \delta_{y_i}(y_\ell)
+\sum_{i\sim_k \ell} \delta_{y_\ell}(y_i)\right)\right\}\,.
\label{eq:knn-cond}
\end{align}
The normalising constant $Z(\beta,k)$ cannot therefore be expressed in closed form. Indeed,
the computation of this constant calls for a summation over $G^n$ terms.

Based on \eqref{eq:knn-cond}, the predictive distribution of a new 
observation $y_{n+1}$ given its covariate $\bx_{n+1}$ and the training sample $(\by,\bX)$ is, for
$g=1,\ldots,G,$
\begin{align*}
&\mathbb{P}(y_{n+1}=g|\bx_{n+1},\by,\bX,\beta,k)\propto\\
&\exp\left\{\beta/k \left(\sum_{\ell\,\sim_k (n+1)} \delta_{g}(y_\ell)+\sum_{(n+1)\sim_k \ell}
\delta_{y_\ell}(g)\right)\right\}\,,
\end{align*}
where 
$$
\sum_{\ell\,\sim_k (n+1)} \delta_{g}(y_\ell)\quad\text{ and }\quad \sum_{(n+1)\sim_k \ell} \delta_{y_\ell}(g)
$$
are the numbers of observations in the training dataset from class $g$ among the $k$ nearest neighbours of $\bx_{n+1}$
and among the observations for which $\bx_{n+1}$ is a $k$-nearest neighbour, respectively
\findeX\end{example}

\section{Monte Carlo Methods}\label{sec:moca}
The generic approach for solving computational problems related with Bayesian analysis is to
use simulation, i.e.~to produce via a computer program a sample from the posterior distribution and to use the 
simulated sample to approximate the procedures of interest. This approach goes under the generic name of Monte
Carlo methods, in reference to the casino of Monaco \citep{metropolis:1987}. Recall that a standard Bayesian 
estimate is the posterior expectation of functions $h(\theta)$ of the parameter,
$$
\mathfrak{I} = \int_\Theta h(\btheta) \pi(\btheta|\by) \,\text{d}\btheta \,.
$$
A formal Monte Carlo algorithm associated with the target $\mathfrak{I}$ proceeds as follows:

\bigskip
\noindent \fbox{
\begin{minipage}{0.46\textwidth}
{\bf Basic Monte Carlo Algorithm}\\
{\sf
For a computing effort $N$
\begin{itemize}
\item[\bf 1)] Set $i=1$,
\item[\bf 2)] Generate independent $\btheta^{(i)}$ from the posterior distribution $\pi(\cdot|\by)$,
\item[\bf 3)] Set $i=i+1$,
\item[\bf 4)] If $i\leq N$, return to {\bf 2)}.
\end{itemize}
}
\end{minipage}
}
\bigskip

The corresponding crude Monte Carlo approximation of $\mathfrak{I}$ is given by:
$$
\widehat{\mathfrak{I}}^\text{MC} = \frac{1}{N}\,\sum_{i=1}^N h\left(\btheta^{(i)}\right)\,.
$$

When the computing effort $N$ grows to infinity, the approximation $\widehat{\mathfrak{I}}^\text{MC}$
converges to $\mathfrak{I}$ and the speed of convergence is $1/\sqrt{N}$ if $h$ is 
square-integrable against $\pi(\btheta|\by)$ \citep{Robert:Casella:2004}. The assessment of
this convergence relies on the Central Limit Theorem, as described in
\citeauthor{robert:casella:2009} (2009, Chapter 4).

\subsection{Importance sampling and resampling}\label{isa}

A generalisation of the basic Monte Carlo algorithm stems from an alternative representation of the
above integral $\mathfrak{I}$, changing both the integrating density and the integrand:
\begin{equation}
\mathfrak{I}=\int_\Theta \frac{h(\btheta) \pi(\btheta|\by)}{g(\btheta)} g(\btheta) \,\text{d}\btheta\,, \label{eq:IS}
\end{equation}
where the support of the posterior distribution $\pi(\cdot|\by)$ is included in the support of $g(\cdot)$.

\bigskip
\noindent \fbox{
\begin{minipage}{0.46\textwidth}
{\bf Importance Sampling Scheme}\\
{\sf
For a computing effort $N$
\begin{itemize}
\item[\bf 1)] Set $i=1$,
\item[\bf 2)] Generate independent $\btheta^{(i)}$ from the importance distribution $g(\cdot)$,
\item[\bf 3)] Calculate the importance weight \\ $\omega^{(i)}=\pi\left(\btheta^{(i)}|\by\right)\big/ g\left(\btheta^{(i)}\right)$,
\item[\bf 4)] Set $i=i+1$,
\item[\bf 5)] If $i\leq N$, return to {\bf 2)}.
\end{itemize}
}
\end{minipage}
}
\bigskip

The corresponding importance sampling approximation of $\mathfrak{I}$ is given by
\begin{equation}\label{eq:Ise}
\widehat{\mathfrak{I}}^\text{IS}_g = \frac{1}{N}\,\sum_{i=1}^N \omega^{(i)} h\left(\btheta^{(i)}\right)\,.
\end{equation}

From a formal perspective, the posterior density $g(\btheta)=\pi(\btheta|\by)$ is a possible (and the most natural) choice for 
the importance function $g(\btheta)$, leading back to the basic Monte Carlo algorithm. However, \eqref{eq:IS} states that a single
integral may be approximated in infinitely many ways. Maybe surprisingly, the choice of the posterior $g(\theta)=\pi(\btheta|\by)$ 
is generaly far from being the most efficient choice of importance function. 
While the representation \eqref{eq:IS} holds in wide generality (the only requirement is that the support of
$\pi(\cdot|\bx)$ should be included in the one of $g(\cdot)$), the choice of $g(\cdot)$ is fundamental to provide
good approximations of $\mathfrak{I}$. Poor choices of $g(\cdot)$ lead to unreliable approximations: for
instance, if 
$$
\ds \int_\Theta h^2(\btheta) \omega^2(\btheta)g(\btheta)\text{d}\btheta
$$
is infinite, the variance of the estimator 
\eqref{eq:Ise} is also infinite \citep[][Chapters 3 and 4]{robert:casella:2009} and then \eqref{eq:Ise}
cannot be used for approximation purposes. 

We stress here that, while Monte Carlo methods do not formaly suffer from the ``curse of dimensionality" in 
the sense that, contrary to numerical methods, the error of the
Monte Carlo estimators is always decreasing in $1/\sqrt N$, notwithstanding the dimension of the parameter
space $\Theta$, the difficulty increases with the dimension $p$ of $\Theta$ in that deriving satisfactory importance sampling distributions
becomes more difficult as $p$ gets larger. As detailed in Section \ref{ISxtend}, a solution for deriving
satisfactory importance functions in large dimensions is to turn to iterative versions of importance sampling.

\begin{figure*}
\begin{center}
 \includegraphics[height=0.33\textwidth]{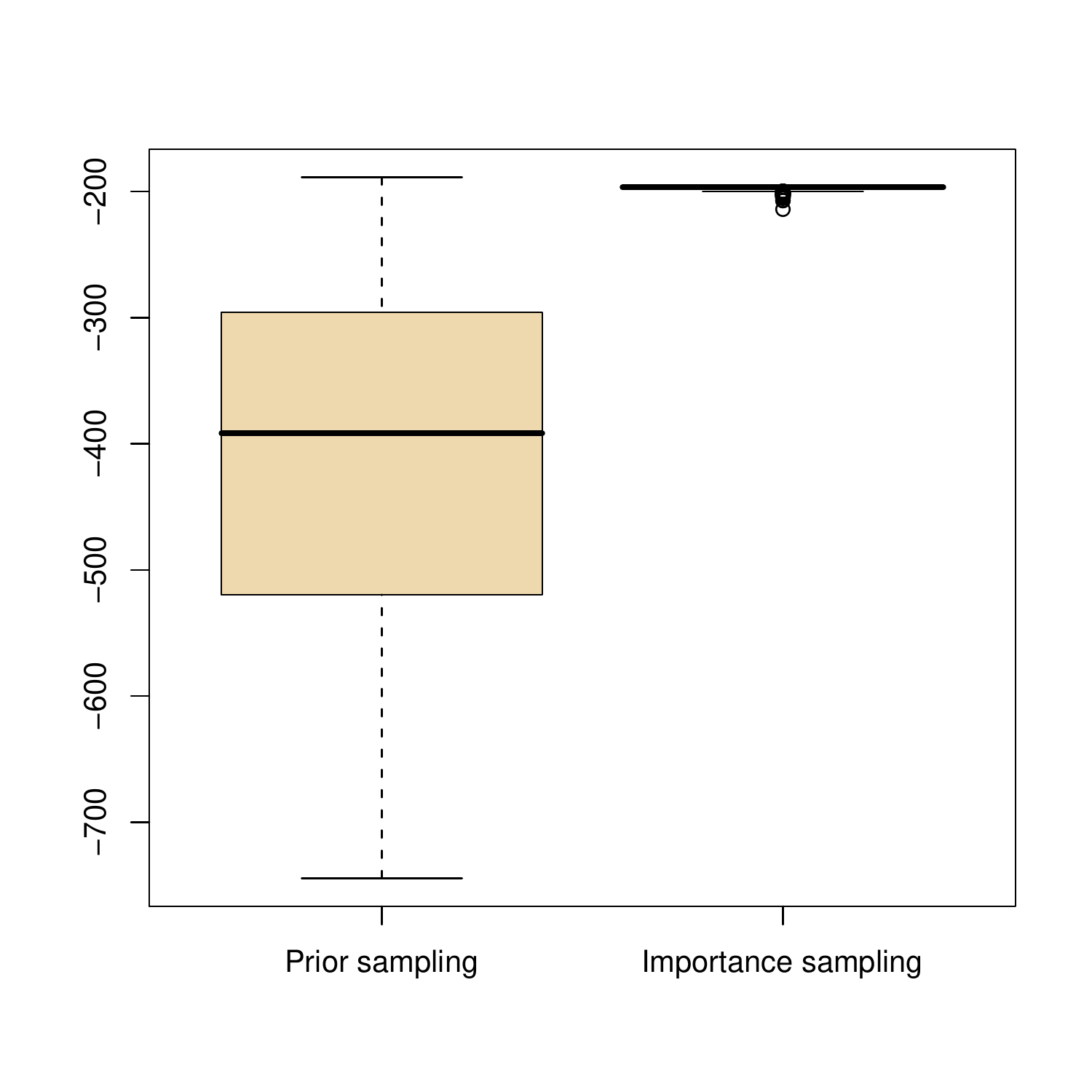}
 \includegraphics[width=0.49\textwidth]{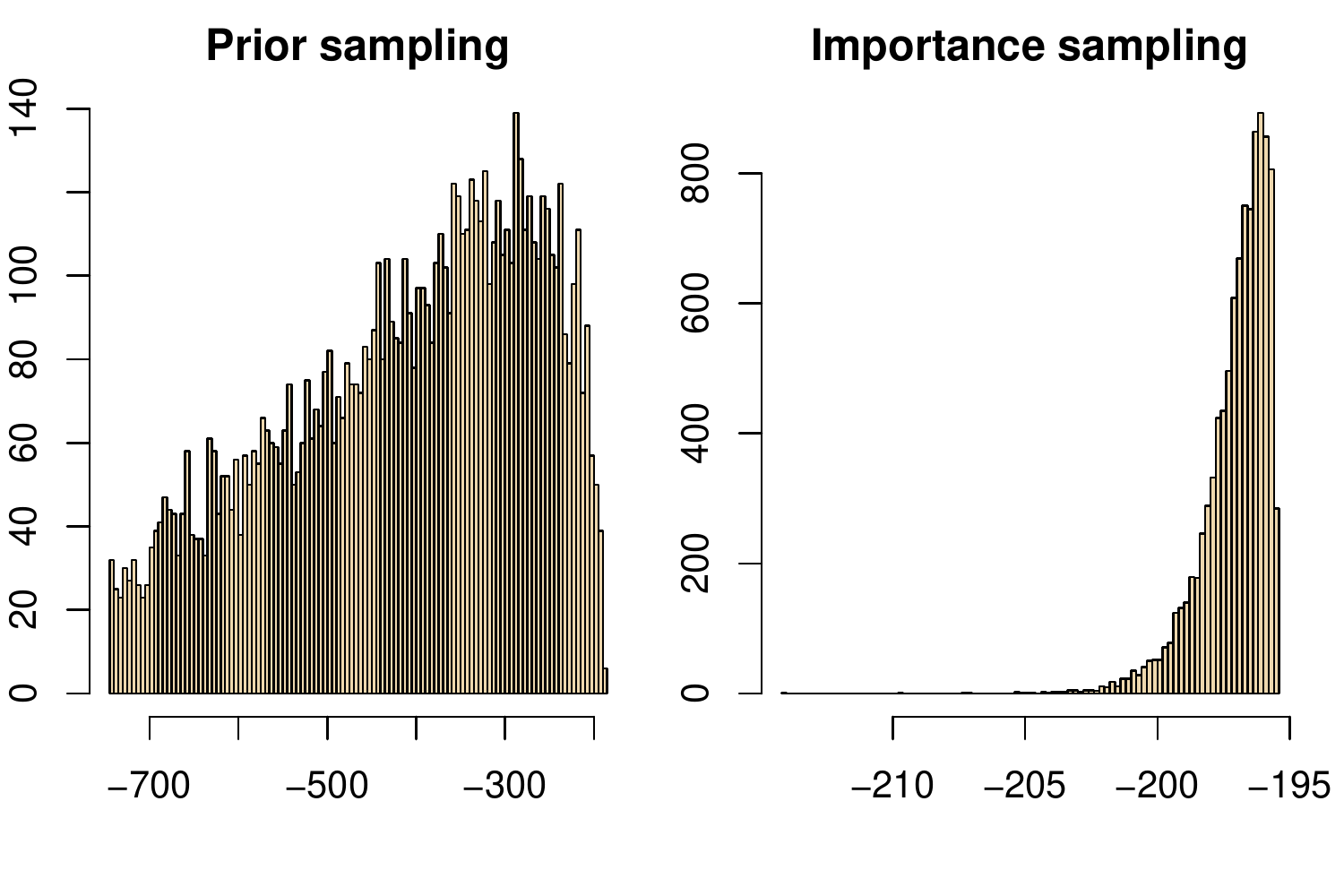}
\end{center}
\caption{\label{fig:compa}
Boxplot and histograms of the logarithms of the importance weights corresponding 
to $10^4$ simulations from the prior distribution {\em (prior sampling)} and 
from the MLE normal approximation {\em (importance sampling)} in the setup of the Pima Indian 
diabetes study of Example \ref{ex:glum}. The graphs for the
prior sampling are excluding the $1690$ zero weights from the representation.
}
\end{figure*}

\begin{example}[Continuation of Example \ref{ex:glum}]\label{ex:glumbit}
\begin{rm}In the case of the probit model, the posterior distribution, proportional to
\eqref{eq:propos} cannot be easily simulated, even though it is bounded from above by
the prior density $\pi(\bbeta|\bX)$.\footnote{This feature means that the accept-reject algorithm
\citep[][Chapter 2]{robert:casella:2009} could formally be used for the simulation of $\pi(\bbeta|\bX,\by)$, 
but the efficiency of this approach would be quite poor.}

In this setting, we propose to use as importance distribution
a normal distribution with mean equal to the maximum likelihood (ML) estimate of $\bbeta$
and with covariance matrix equal to the estimated covariance matrix of the ML estimate.
While, in general, those normal distributions provide crude approximations to the posterior
distributions, the specific case of the probit model shows this is an exceptionally
good approximation to the posterior. For instance, if we compare the weights resulting from
using this normal distribution with the weights resulting from using the prior distribution
as importance function, the range of the former weights is much more concentrated than for
the later weights, as shown by Figure \ref{fig:compa}. (Note that, due to the missing normalising
constant in $\pi(\bbeta|\bX,\by)$, the weights are computed with the product
\eqref{eq:propos} as the target function.)
\findeX\end{example}

As noted in the above example, a common feature of Bayesian integration settings is that the normalising constant of
the posterior distribution, $m(\by)$, cannot be computed in closed form. In that case,
$\omega^{(i)}$ and $\widehat{\mathfrak{I}}^\text{IS}_g$ cannot be used and they are replaced by the unormalised version
$$
\omega^{(i)} = m(\by) \pi(\btheta^{(i)}|\by) / g(\btheta^{(i)})
$$
and by the self-normalized version
$$
\widehat{\mathfrak{I}}^\text{SNIS}_g = \sum_{i=1}^N \omega^{(i)} h\left(\btheta^{(i)}\right)
\bigg/ \sum_{i=1}^N \omega^{(i)}\,,
$$
respectively. The self-normalized $\widehat{\mathfrak{I}}^\text{SNIS}_g$
also converges to $\mathfrak{I}$ since $\sum_{i=1}^N \omega^{(i)}$ converges to 
the normalising constant $m(\by)$. 
The weights $(i=1,\ldots,T)$
$$
\overline\omega^{(i)} = \omega^{(i)} \big/ \sum_{j=1}^N \omega^{(j)}
$$
are then called {\em normalised weights} and, since they sum up to one, they induce a
probability distribution on the sample of $\theta^{(i)}$'s. When chosing an importance function, 
the adequation with the posterior distribution needs to get higher
as the dimension $p$ increases. Otherwise, very few weights $\overline\omega^{(i)}$
are different from $0$ and even the largest weight, which is then close to $1$, may correspond to an unlikely 
value for the posterior distribution, its closeness to $1$ being then an artifact of the renormalisation 
and not an indicator of its worth. A related measure of performance of the importance function is given by the
{\em effective sample size}

$$
\text{ESS}_N = 1\bigg/ {\sum_{i=1}^N \left(\overline\omega^{(i)}\right)^2}\,.
$$
For a uniformly weighted sample, $\text{ESS}_N$ is equal to $N$, while, for a completely degenerated sample 
where all importance weights but one are zero, $\text{ESS}_N$ is equal to $1$. The effective sample size thus 
evaluates the size of the iid sample equivalent to the weighted sample and allows for a direct comparison of samplers.

\begin{example}[Continuation of Example \ref{ex:glumbit}]\label{ex:ESSbit}\begin{rm}
For the two schemes tested in the probit model of Example \ref{ex:glumbit}, using the same number $N=10,000$ of simulations,
the effective sample sizes are $T_1=6291.45$ and $T_2=9.77$ for the ML based and prior normal importance functions, respectively.
\findeX\end{example}

While importance sampling is primarily an integral approximation methodology, it can also be 
used for simulation purposes, via the {\em sampling importance resampling} (SIR) methodology
of \cite{Rubin:1988}. Given a weighted sample $(\btheta^{(1)},\overline\omega^{(1)}),\ldots,(\btheta^{(N)},\overline\omega^{(N)})$
simulated from $g(\cdot)$, it is possible to derive a sample approximately distributed from the target distribution $\pi(\cdot|\bx)$,
$\tilde \btheta^{(1)},\ldots,\tilde \btheta^{(M)}$, by resampling from the instrumental sample $\btheta^{(1)},\ldots,\btheta^{(N)}$ 
using the importance weights, that is,
$$
\tilde \btheta^{(i)}= \btheta^{(J_i)}\,,\quad\quad 1\le i\le M\,,
$$
where the random variables $J_1,\ldots,J_M$ are distributed as
$$
\mathbb{P}\left(J_l=i\left|\btheta^{(1)},\ldots,\btheta^{(N)}\right.\right) = \overline\omega^{(i)}
$$
\citep[see, e.g.,][Chapter 3]{robert:casella:2009}.

\subsection{Sequential importance sampling}\label{ISxtend}
In general, importance sampling techniques require a rather careful tuning
to be of any use, especially in large dimensions. While MCMC methods (Section \ref{sec:mcmc})
are a ready-made solution to this problem, given that they can break the global distribution into 
distributions with smaller dimensions, the recent literature has seen an extension of importance 
sampling that adaptively calibrates some importance functions towards more similarity with the target density 
\citep{Cappe:Guillin:Marin:Robert:2003,delmoral:doucet:jasra:2006,Douc:Guillin:Marin:Robert:2005,cappe:douc:guillin:marin:robert:2007}.

The method is called sequential Monte Carlo (SMC) because it evolves along a time axis either through
the target---as in regular sequential statistical problems---or through the importance function, and also 
population Monte Carlo (PMC), following \cite{Iba:2000}, because it produces {\em populations} rather than 
points.\footnote{This simulated population is then used to devise new and hopefully improved importance (or proposals) functions.} 
Although the idea has connections with the earlier particle filter literature
\citep{gordon:salmon:smith:1993,doucet:defreitas:gordon:2001,cappe:moulines:ryden:2004}, the main
principle of this method is to build a sequence of increasingly better---against a performance criterion
that may be the entropy divergence from the target distribution or the variance of the corresponding
estimator of a fixed integral---proposal distributions
through a sequence of simulated samples (which thus behave like populations). 
Given that the validation of the technique is still based 
on sampling importance resampling principles, the resulting dependence on past samples can be arbitrarily complex, 
{\em while} the approximation to the target remains valid (unbiased) at {\em each iteration} and 
while it does not require asymptotic convergence as MCMC methods do (see Section \ref{sec:mcmc}). 
A very recent connection between both approaches can be found in \cite{andrieu:doucet:holenstein:2010} and the discussion therein.

While the following algorithm does appear as a repeated (or sequential) sampling importance 
resampling algorithm \citep{Rubin:1988}, the major update is the open choice of $q_{it}$ 
in the first step, since $q_{it}$ can depend on all past simulated samples as well as on the
index of the currently simulated value. For instance, in \cite{cappe:douc:guillin:marin:robert:2007},
mixtures of standard kernels are used with an update of the weights and of the parameters of those
kernels at each iteration in such a way that the entropy distance 
of the corresponding importance sampling estimator to the target are decreasing from one 
iteration to the other.

\bigskip
\noindent \fbox{
\begin{minipage}{0.46\textwidth}
{\bf General Population Monte Carlo Algorithm}\\
{\sf
For a computing effort $N$
\begin{itemize}
\item[\bf 1)] Generate $(\btheta_{i,0})_{1 \leq i\leq N} \stackrel{\text{iid}}{\sim} q_0$ and compute $\ds \omega_{i,0}=\pi(\btheta_{i,0}|\by)/q_0(\btheta_{i,0})$,
\item[\bf 2)] Generate $(J_{i,0})_{1\leq i\leq N} \stackrel{\text{iid}}{\sim} \mathcal{M}(1,(\overline \omega_{i,0})_{1\leq i\leq N})$ and set $\tilde \btheta_{i,0}=\btheta_{J_{i,0},0}$ $(1\le i\le N)$,
\item[\bf 3)] Set $t=1$,
\item[\bf 4)] Conditionally on past $\btheta_{i,j}$'s and $\tilde \btheta_{i,j}$'s, generate independently $\btheta_{i,t}\sim q_{i,t}$ and compute 
$\ds \omega_{i,t}=\pi(\btheta_{i,t}|\by)/q_{i,t}(\btheta_{i,t})$,
\item[\bf 4)] Generate
$
(J_{i,t})_{1\leq i\leq N} \stackrel{\text{iid}}{\sim} \mathcal{M}(1,(\overline \omega_{i,t})_{1\leq i\leq N})
$
and set $\tilde \btheta_{i,t}=\btheta_{J_{i,t},t}$ $(1\le i\le N)$,
\item[\bf 5)] Set $t=t+1$,
\item[\bf 6)] If $t\leq N$ return to {\bf 4)}.
\end{itemize}
}
\end{minipage}
}

\bigskip
In this representation, while the choice of $q_{it}$ is completely open, a convenient case is
when the $\btheta_{i,t}$'s are simulated either from a non-parametric kernel-like proposal of the form 
$$
\sum_{j=1}^n \varrho_{j,t-1}\,K_t(\tilde\btheta_{j,t-1},\btheta)\,,
$$
where $K_t$ is a Markov kernel modified at each iteration \citep{Douc:Guillin:Marin:Robert:2007b} or from a 
mixture of the form
$$
\sum_{j=1}^n \varrho_{j,t-1}\,g_j(\tilde\btheta_{j,t-1}|\xi_{j,t-1})\,,
$$
where $g_j$ is a standard distribution from an exponential family parameterised by $\xi$, both parameters
and weights being updated at each iteration \citep{cappe:douc:guillin:marin:robert:2007}. An illustration
of the performances of this PMC algorithm for a cosmological target is given in \cite{wraith-2009-80}, while
an ABC extension has been introduced by \cite{beaumont:cornuet:marin:robert:2009}.

Since PMC produces at each iteration a valid approximation to the target distribution, the populations $\tilde\btheta_{i,t}$
produced at each of those iterations should not be dismissed for approximation purposes. \cite{cornuet:marin:mira:robert:2009}
have developped a nearly optimal strategy recycling all past simulations, based on the multiple mixture technique of
\cite{owen:zhou:2000} and called {\em adaptive multiple importance sampling} (AMIS). 

\subsection{Approximations of the Bayes factor}\label{Bprox}

As already explained above, when testing 
for an null hypothesis (or a model) $H_0:\btheta\in\Theta_0$ against the alternative hypothesis
(or the alternative model) $H_1:\btheta\in\Theta_1$, the Bayes factor is defined by
$$
B_{01}(\by)={\displaystyle{\int_{\Theta_0} f_0(\by|\btheta_0) \pi_0(\btheta_0) \text{d}\btheta_0} }\bigg/
$$
$$
{\displaystyle{\int_{\Theta_1} f_1(\by|\btheta_1) \pi_1(\btheta_1) \text{d}\btheta_1} }\,.
$$
The computation of Monte Carlo approximations of the Bayes factor \eqref{Baf} has 
undergone rapid changes in the last decade as illustrated by the book of \cite{Chen:Shao:Ibrahim:2000}
and the recent survey of \cite{robert:marin:2010}.
We assume here that the prior distributions under both the null and the alternative hypotheses are 
proper, as, typically they should be. (In the case of common nuisance parameters, a common improper prior
measure can be used on those, see \cite{berger:pericchi:varshavsky:1998,marin:robert:2007}. This 
complicates the computational aspect, as some methods like crude Monte Carlo cannot be used at all, while others
are more prone to suffer from infinite variance.)
In that setting, the most elementary approximation to $B_{01}(\by)$ consists in using a ratio of two
standard Monte Carlo approximations based on simulations from the corresponding priors. Indeed, for 
$i=0,1$:
$$
\int_{\Theta_i} f_i(\by|\btheta) \pi_i(\btheta) \text{d}\btheta=\mathbb{E}_{\pi_i}\left[f(\by|\btheta)\right]\,.
$$
Then, if $\btheta_{0,1},\ldots,\btheta_{0,n_0}$ and $\btheta_{1,1},\ldots,\btheta_{1,n_1}$
are two independent samples generated from the prior distributions $\pi_0$ and $\pi_1$, respectively, 
$$
\frac{n_0^{-1}\sum_{j=1}^{n_0}f_0(\by|\btheta_{0,j})}{n_1^{-1}\sum_{j=1}^{n_1}f_1(\by|\btheta_{1,j})}
$$
is a strongly consistent estimator of $B_{01}(\by)$.

Defining two importance distributions with densities $g_0$ and $g_1$, with 
the same supports as $\pi_0$ and $\pi_1$, respectively, we have:
\begin{align*}
B_{01}(\by)=&{\mathbb{E}_{g_0}\left[f_0(\by|\btheta)\pi_0(\btheta)\big/g_0(\btheta)\right]}
\bigg/\\
&{\mathbb{E}_{g_1}\left[f_1(\by|\btheta)\pi_1(\btheta)\big/g_1(\btheta)\right]}\,.
\end{align*}
Therefore, given two independent samples generated from distributions $g_0$ and $g_1$,
respectively, $\btheta_{0,1},\ldots,\btheta_{0,n_0}$ and $\btheta_{1,1},\ldots,\btheta_{1,n_1}$,
the corresponding importance sampling estimate of $B_{01}(\by)$ is
$$
\dfrac{n_0^{-1} \sum_{j=1}^{n_0} f_0(\by|\btheta_{0,j}) \pi_0(\btheta_{0,j})/g_0(\btheta_{0,j})}
{n_1^{-1} \sum_{j=1}^{n_1} f_1(\by|\btheta_{1,j}) \pi_1(\btheta_{1,j})/g_1(\btheta_{1,j})}\,.
$$
Compared with the standard Monte Carlo approximation above, this approach offers
the advantage of opening the choice of the representation in that 
it is possible to pick importance distributions $g_0$ and $g_1$ that lead
to a significant reduction in the variance of the importance sampling estimate.

In the special case when the parameter spaces of both models under comparison are  
identical, i.e.~$\Theta_0=\Theta_1$, a bridge sampling approach \citep{Meng:Wong:1996} is based on the general representation
\begin{eqnarray*}
B_{01}(\by) & = & {\ds \int f_0(\by|\btheta) \pi_0(\btheta) \alpha(\btheta) {\pi}_1(\btheta|y) \text{d}\btheta }\bigg/ \\
& & {\ds \int f_1(\by|\btheta) \pi_1(\btheta) \alpha(\btheta) {\pi}_0(\btheta|\by) \text{d}\btheta } \\  
& \approx & 
\frac{\ds {n_1}^{-1} \sum_{j=1}^{n_1} f_0(\by|\btheta_{1,j}) \pi_0(\btheta_{1,j}) \alpha(\btheta_{1,j})}
{\ds {n_0}^{-1} \sum_{j=1}^{n_0} f_1(\by|\btheta_{0,j}) \pi_1(\btheta_{0,j}) \alpha(\btheta_{0,j})}
\end{eqnarray*}
where $\btheta_{0,1},\ldots,\btheta_{0,n_0}$ and $\btheta_{1,1},\ldots,\btheta_{1,n_1}$
are two independent samples coming from the posterior distributions $\pi_0(\btheta|\by)$
and $\pi_1(\btheta|\by)$, respectively. That applies for any positive function $\alpha$ such that 
the upper integral exists. Some choices of $\alpha$ can lead to very poor performances of the method in connection 
with the harmonic mean approach (see below), but there
exists a quasi-optimal solution, as provided by \cite{gelman:meng:1998}:
$$
{\alpha^\star(\by) \propto \dfrac{1}{n_0{\pi}_0(\btheta|\by) + n_1  {\pi}_1(\btheta|\by)}} \,.
$$
This optimum cannot be used {\em per se}, since it requires the normalising constants of
both $\pi_0(\btheta|\by)$ and $\pi_1(\btheta|\by)$. As suggested by \cite{gelman:meng:1998},
an approximate but practical version uses iterative versions of $\alpha^\star$, the current 
approximation of $\alpha^\star$ being used to produce a new bridge sampling approximation of 
$B_{01}(\by)$, which in its turn is used to set a
new approximation of $\alpha^\star$. Note that this solution recycles simulations from both posteriors, 
which is quite appropriate since one model is selected via the Bayes factor, 
instead of using an importance weighted sample common to
both approximations. We will see below an alternative representation of the bridge factor 
that bypasses this difficulty (if difficulty there is!).

Those derivations are however restricted to the case when both models have the same complexity and thus 
they do not apply to embedded models, when $\Theta_0\subset\Theta_1$ in such a way that 
$\btheta_1=(\btheta,\psi)$, i.e.~when the submodel corresponds to a specific value $\psi_0$ of $\psi$:
$f_0(\by|\btheta)=f_1(\by|\btheta,\psi_0)$.

The extension of the most advanced bridge sampling strategies to such cases 
requires the introduction of a {\em pseudo-posterior density,} $\omega(\psi|\btheta,\by)$, on the
parameter that does not appear in the embedded model,
in order to reconstitute the equivalence between both parameter spaces. Indeed, 
if we augment $\pi_0(\btheta|\by)$ with $\omega(\psi|\btheta,\by)$, we obtain a joint 
distribution with density $\pi_0(\btheta|\by)\times\omega(\psi|\btheta,\by)$ on $\Theta_1$.
The Bayes factor $B_{01}(\by)$  can then be expressed as

{\footnotesize{\begin{equation}\label{eq:psudo}
\dfrac{\ds \int_{\Theta_1} f_1(\by|\btheta,\psi_0) \pi_0(\btheta)\alpha(\btheta,\psi)\pi_1(\btheta,\psi|\by) 
\text{d}\btheta\omega(\psi|\btheta,\by)\,\text{d}\psi}
{\ds \int_{\Theta_1} f_1(\by|\btheta,\psi) \pi_1(\btheta,\psi) \alpha(\btheta,\psi)\pi_0(\btheta|\by) 
\omega(\psi|\btheta,\by) \text{d}\btheta \,\text{d}\psi}\,,   
\end{equation}
}}

\noindent because it is clearly independent from the choice of both $\alpha(\btheta,\psi)$ and $\omega(\psi|\btheta,\by)$. 
Obviously, the performances of the approximation

{\footnotesize{$$
\dfrac{\ds (n_1)^{-1} \sum_{j=1}^{n_1} f_1(\by|\btheta_{1,j},\psi_0) 
\pi_0(\btheta_{1,j}) \omega(\psi_{1,j}|\btheta_{1,j},\by)\alpha(\btheta_{1,j},\psi_{1,j})}
{\ds (n_0)^{-1} \sum_{j=1}^{n_0} f_1(\by|\btheta_{0,j},\psi_{0,j}) \pi_1(\btheta_{0,j},
\psi_{0,j})  \alpha(\btheta_{0,j},\psi_{0,j})}\,,
$$
}}

\noindent where $(\btheta_{0,1},\psi_{0,1}),\ldots,(\btheta_{0,n_0},\psi_{0,n_0})$ and 
$(\btheta_{1,1},\psi_{1,1}),\ldots,(\btheta_{1,n_1},\psi_{1,n_1})$
are two independent samples generated from distributions $\pi_0(\btheta|\by)\times\omega(\psi|\btheta,\by)$
and $\pi_1(\theta,\psi|\by)$, respectively, do depend on this completion by the pseudo-posterior
as well as on the function $\alpha(\btheta,\psi)$.
\cite{Chen:Shao:Ibrahim:2000} establish that the asymptotically optimal choice for 
$\omega(\psi|\btheta,\by)$ is the obvious one, namely
$$
\omega(\psi|\btheta,\by) = \pi_1(\psi|\btheta,\by) \,,
$$
which most often is unavailable in closed form (especially when considering that 
the normalising constant of $\omega(\psi|\btheta,\by)$ is required in \eqref{eq:psudo}).

Another approach to approximating the marginal likelihood is based on harmonic means. If $\btheta_{i,j}\sim\pi_i(\cdot)$
$(i=1,2,\,j=1,\ldots,N)$, the prior distribution, then
$$
\frac{1}{N}\,\sum_{j=1}^N \frac{1}{f_i(\by|\btheta_{i,j})}
$$
is an unbiased estimator of $1/m_i(\by)$ \citep{newton:raftery:1994}. This generic harmonic mean is too often associated
with an infinite variance to ever be recommended \citep{neal:1994}, but the representation \citep{gelfand:dey:1994} $(i=0,1)$
\begin{align*}
\mathbb{E}_{\pi_i}\left[\left.\frac{\varphi_i(\btheta) }{\pi_i(\btheta)f_i(\by|\btheta)}\right| \by \right]
&= \int \frac{\varphi_i(\btheta) \pi_i(\btheta)f_i(\by|\btheta)}
{\pi_i(\btheta)f_i(\by|\btheta) m_i(\by)}\,\text{d}\btheta\\ &= \frac{1}{m_i(\by)}
\end{align*}
holds, no matter what the density $\varphi_i$ is, provided $\varphi_i(\btheta_i)=0$ when $\pi_i(\btheta_i)f_i(\by|\btheta_i)=0$.
This representation is remarkable in that it allows for a direct processing of Monte Carlo (or MCMC) output from the posterior distribution
$\pi_i(\btheta_i|\by)$. As with importance sampling approximations, the variability of the corresponding estimator of $B_{01}(\by)$ will
be small if the distributions $\varphi_i$ ($i=0,1$) are close to the corresponding posterior distributions. However,
as opposed to usual importance sampling constraints, the density
$\varphi_i$ must have lighter---rather than fatter---tails than $\pi_i(\cdot)f_i(\by|\cdot)$
for the approximation of the marginal $m_i(x)$
$$
\left[ N^{-1}\,\sum_{j=1}^N \frac{\varphi_i(\btheta_{i,j})}{\pi_i(\btheta_{i,j}) f_i(\by|\btheta_{i,j})}\right]^{-1}\,,
$$
when $\btheta_{i,j}\sim\pi_i(\btheta|\by)$,
to enjoy finite variance. For instance, using $\varphi_i$'s with constrained supports derived from a Monte Carlo sample,
like the convex hull of the simulations corresponding to the $10\%$ or to the $25\%$ HPD regions---that again is easily derived
from the simulations---is both completely appropriate and implementable \citep{robert:wraith:2009}.

\begin{figure}
 \includegraphics[height=5cm,width=7cm]{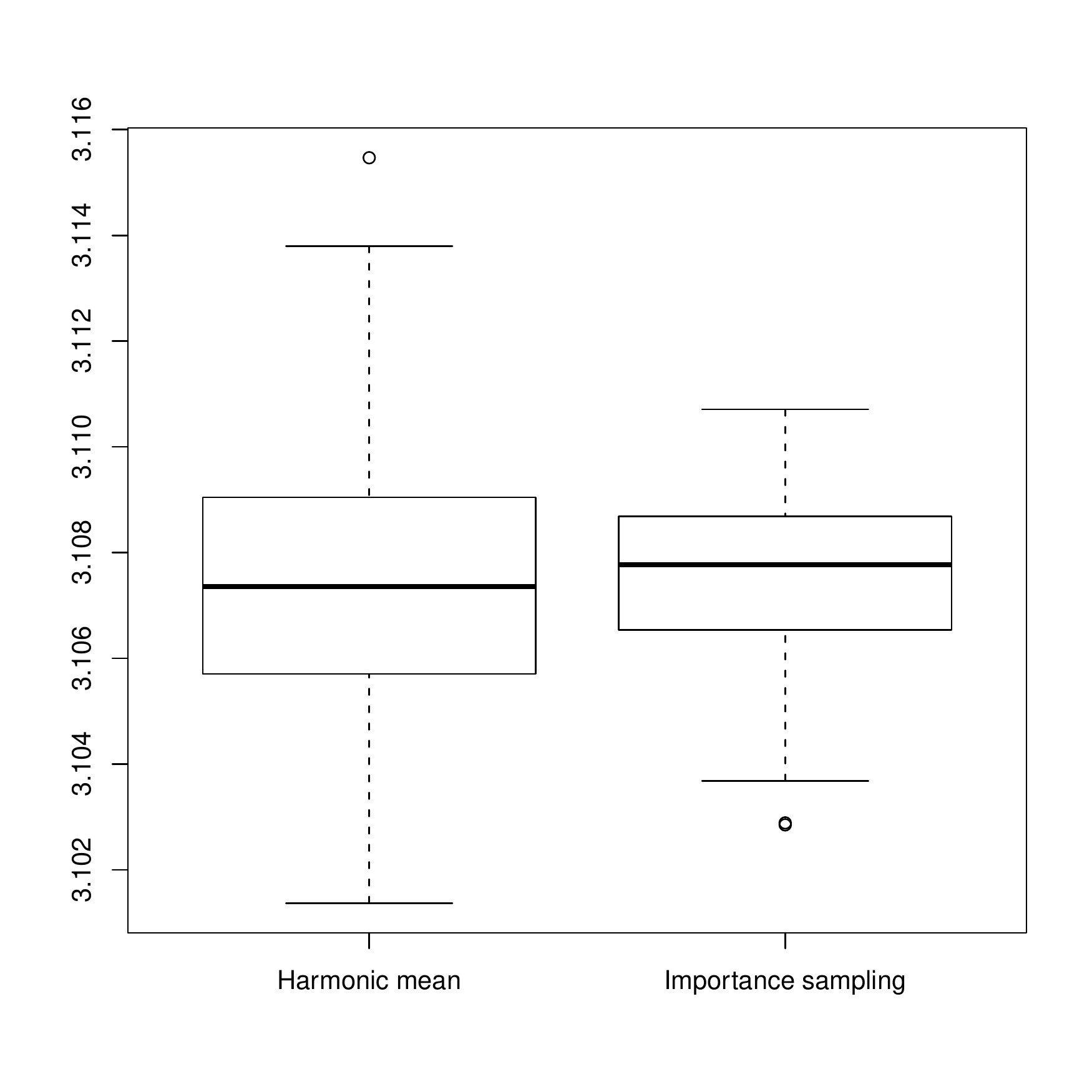} 
\caption{\label{fig:bfhm}
Monte Carlo experiment comparing the variability of the approximations to the Bayes factor $B_{10}(\by)$
based on harmonic mean and importance sampling for the Pima Indian diabetes study of Example \ref{ex:glum}. 
The boxplots are obtained for $100$ replications of $20,000$ simulations from the normal importance sampling
distribution.  {\em (Source: \citealp{robert:marin:2010})}.}
\end{figure}

\begin{example}[Continuation of Example \ref{ex:glumbit}]\label{ex:glumic}
\begin{rm}In the case of the probit model, if we use as distributions $\varphi_i$ the
normal distributions with means equal to the ML estimates and covariance matrices equal to the estimated covariance
matrices of the ML estimates, the results of \cite{robert:marin:2010}, obtained over $100$ replications
with $N=20,000$ simulations each are reproduced in Figure
\ref{fig:bfhm}. They compare both approaches---harmonic mean and importance sampling---to the approximation of the Bayes factor testing
for the significance of the {\sf ped} covariate and show a very clear proximity between both importance solutions in this
special case, even though the importance sampling estimate is much faster to compute. Simulation from the posterior distribution
is obtained by an MCMC algorithm described in Section \ref{sec:mcmc}.
\findeX\end{example}

A final approach to the approximation of Bayes factors that is worth exploring is 
Chib's (1995) method. First, it is a direct application of
Bayes' theorem: given $\by\sim f_i(\by|\btheta)$, we have that
$$
m_i(\by) = \frac{f_i(\by|\btheta)\,\pi_i(\btheta)}{\pi_i(\btheta|\by)}\,,
$$
for all $\btheta$'s (since both the lhs and the rhs of this equality are constant in $\btheta$). Therefore, if an
arbitrary value $\btheta^*$, is selected and if a good approximation to $\pi_i(\btheta^*|\by)$
is available, denoted $\hat\pi_i(\btheta^*|\by)$, Chib's (\citeyear{chib:1995}) approximation to the marginal
likelihood (and hence to the Bayes factor) is
\begin{equation}\label{eq:chib}
m_i(\by) = \frac{f_i(\by|\btheta^*)\,\pi_i(\btheta^*)}{\hat\pi_i(\btheta^*|\by)}\,.
\end{equation}
In a general setting, $\hat{\pi}_i(\theta^*|\by)$ may be the normal approximation based on the MLE, already
used in the importance sampling, bridge sampling and harmonic mean solutions, but this is unlikely to be accurate in a
general framework.
A second solution is to use a nonparametric approximation based on a preliminary MCMC sample, even though the accuracy may
also suffer in large dimensions.  In the special setting of latent variables models introduced in Section \ref{sec:watsup},
Chib's (1995) approximation is particularly attractive as there exists
a natural approximation to $\pi_k(\btheta|\by)$, based on the Rao--Blackwell
\citep{gelfand:smith:1990} estimate
$$
\hat\pi_k(\btheta^*|\by) = \frac{1}{N}\,\sum_{j=1}^N \pi_k(\btheta^*|\by,\bz_j)\,,
$$
where the $z_j$'s are the latent variables simulated by the MCMC sampler.
The estimate $\hat\pi_k(\theta^*|\by)$
is indeed a parametric unbiased approximation of $\pi_k(\theta^*|\by)$ that converges with rate $\text{O}(1/\sqrt{N})$.
It obviously requires the full conditional density $\pi_k(\btheta^*|\by,\bz)$ to be
available in closed form (constant included) but, for instance, this is the case for the probit model of Example
\ref{ex:glum}.
 
\begin{figure}
 \centerline{\includegraphics[width=7cm]{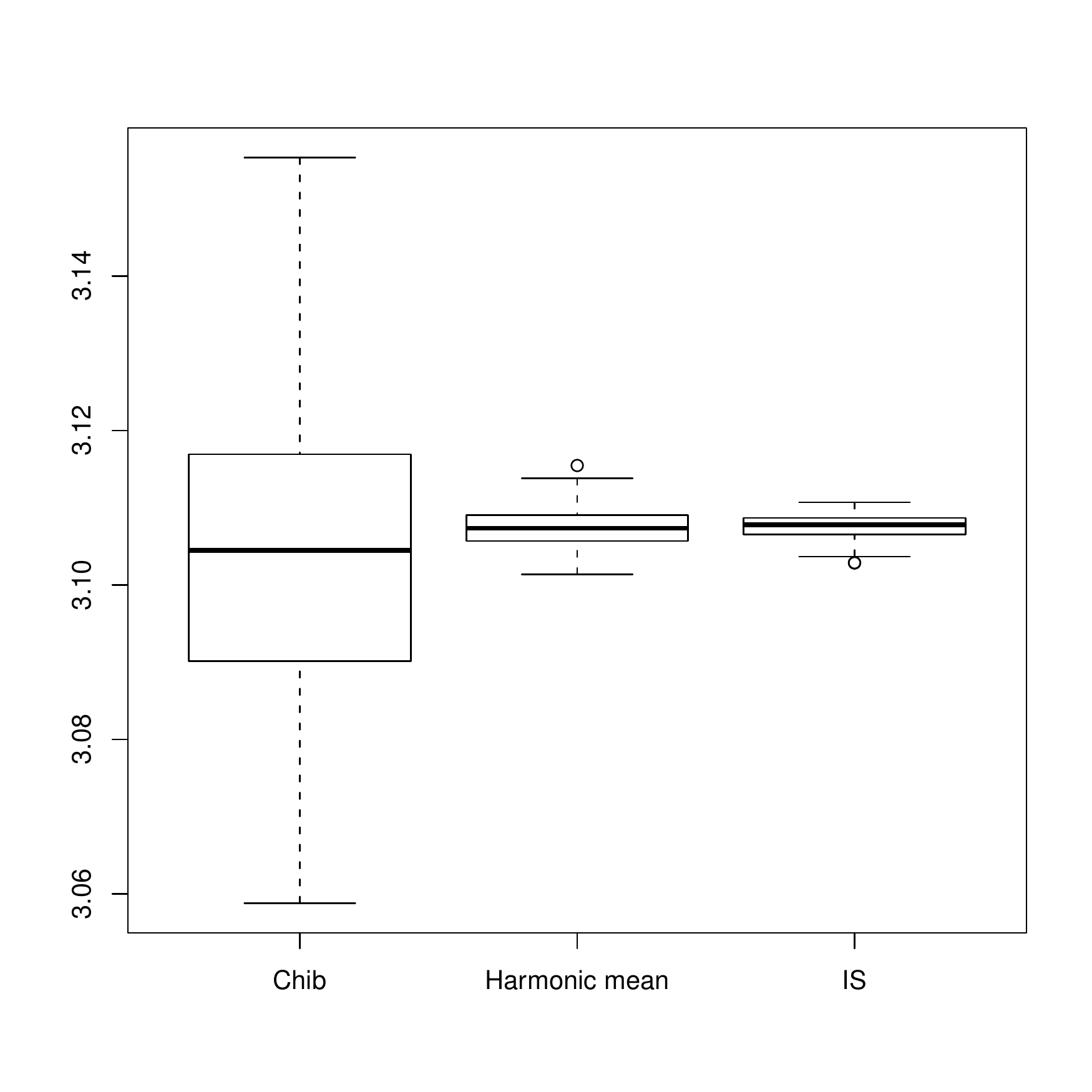}}
\caption{\label{fig:bfchi} Monte Carlo experiment comparing the variability of the approximations to the Bayes factor $B_{10}(\by)$
based on Chib's representation and on harmonic mean and importance sampling for the Pima Indian diabetes study of Example \ref{ex:glum}.
The boxplots are obtained for $100$ replications of $20,000$ simulations from the posterior and the importance sampling
distributions, respectively.  {\em (Source: \citealp{robert:marin:2010})}.}
\end{figure}

\begin{example}[Continuation of Example \ref{ex:glumic}]\label{ex:glumib}\begin{rm}
Figure \ref{fig:bfchi} reproduces the results of \cite{robert:marin:2010} obtained for $100$ replications 
of Chib's approximations of $B_{01}(\by)$ for the same test as in Example \ref{ex:glumic}
with $N=20,000$ simulations for each approximation of $m_i(\by)$ ($i=0,1$). While Chib's method is usually
very reliable and dominates importance sampling, the incredibly good approximation provided by the asymptotic normal
distribution implies that, in this highly special case, Chib's method is dominated by both the importance sampling and the
harmonic mean estimates.
\findeX\end{example}

While the methods presented above cannot ranked in a fixed order for all types of problems, the conclusion of 
\citealp{robert:marin:2010} is worth repeating here. In cases when a good approximation $g(\cdot)$ to the true posterior
distribution of a model is available, it should be used in a regular importance sampling evaluation of the marginal 
likelihood. Given that this good fit rarely occurs in complex and new settings, more generic solutions like Chib's (1995)
should be used, whenever available. (When used with a bounded support on the $\varphi_i$'s, the harmonic mean approximation
can be considered as a generic method.) At last, when faced with a large number or even an infinity of models to compare, 
the only available solution is to use model jump techniques like reversible jump MCMC \citep{green:1995}.

\section{Markov chain Monte Carlo methods}\label{sec:mcmc}

\subsection{Basics}
Given the difficulties involved in constructing an efficient importance function in complex setting,
Markov chain Monte Carlo (MCMC) methods \citep{gelfand:smith:1990,Robert:Casella:2004,robert:casella:2009,marin:robert:2007}
try to overcome some of the limitations of regular Monte Carlo methods (particularly
dimension-wise) by simulating a Markov 
chain with stationary (and limiting) distribution the target distribution.\footnote{The theoretical
foundations of MCMC algorithms are both sound and simple: as stressed by \cite{tierney:1994} and 
\cite{mengersen:tweedie:1996}, the existence of a stationary distribution almost immediately validates
the principle of a simulation algorithm based on a Markov kernel.} There exist 
fairly generic ways of producing such chains, including the Metropolis--Hastings and Gibbs
algorithms defined below. Besides the fact that stationarity of the target
distribution is enough to justify a simulation method by Markov chain generation,
the idea at the core of MCMC algorithms is that {\em local} exploration, when properly
weighted, can lead to a valid ({\em global\/}) representation of the distribution of interest. This 
includes for instance using only component-wise (and hence small-dimensional) simulations---that 
escape (to some extent) the curse of dimensionality---as in the Gibbs sampler. 

This very short introduction may give the impression that MCMC simulation is only superficially different from other Monte Carlo
methods. When compared with alternatives such as importance sampling, MCMC methods differ on two issues:
\begin{enumerate}
\item the output $(\theta^{(t)})$ of an MCMC algorithm is only asymptotically distributed from the target distribution. While this 
usually is irrelevant, in that the $\theta^{(t)}$'s are very quickly distributed from that target, it may also happen that the
algorithm fails to converge in the prescribed number of iterations and thus that the resulting estimation is biased;
\item the sequence $(\theta^{(t)})$ being a Markov chain, the $\theta^{(t)}$'s are correlated and therefore this modifies the evaluation
of the asymptotic variance as well as the effective sample size associated with the output.
\end{enumerate}
We also note here that trying to produce an iid sequence out of a MCMC method is highly inefficient and thus not recommended.

\subsection{Metropolis--Hastings algorithm}
The Metropolis--Hastings algorithm truly is {\em the} generic MCMC method in that
it offers a straightforward and universal solution to the problem of simulating 
from an arbitrary\footnote{The only restriction is that this function is known up to
a normalising constant.} posterior distribution $\pi(\btheta|\bx)\propto f(\by|\btheta)\,\pi(\btheta)$: starting from
an arbitrary point $\btheta_0$, the corresponding Markov chain explores the
surface of this posterior distribution using an internal Markov kernel (proposal) $q(\btheta|\btheta^{(t-1)})$
that progressively visits the whole range of the possible values of $\btheta$. This internal Markov
kernel should be irreducible with respect to the target distribution (that is, the Markov chain associate
whith the proposal $q(\cdot|\cdot)$ should be able to visit the whole support of the target distribution).
The reason why the resulting chain does converge to the target distribution despite the arbitrary choice
of $q(\btheta|\btheta^{(t-1)})$ is that the proposed values are sometimes rejected by a step that
relates with the accept-reject algorithm.

\bigskip
\noindent \fbox{
\begin{minipage}{0.46\textwidth}
{\bf Metropolis--Hastings Algorithm}\\
{\sf
For a computing effort $N$
\begin{itemize}
\item[\bf 1)] Choose $\btheta^{(0)}$,
\item[\bf 2)] Set $t=1$,
\item[\bf 3)] Generate $\btheta'$ from $q(\cdot|\btheta^{(t-1)})$,
\item[\bf 4)] Generate $u$ from $\mathcal{U}_{[0,1]}$,
\item[\bf 5)] If $\ds u \leq \frac{\pi(\btheta')f(\by|\btheta')q(\btheta^{(t-1)}|\btheta')}{\pi(\btheta^{(t-1)}f(\by|\btheta^{(t-1)})q(\btheta'|\btheta^{(t-1)})}$, \\
set $\btheta^{(t)}=\btheta'$ else $\btheta^{(t)}=\btheta^{(t-1)}$,
\item[\bf 6)] Set $t=t+1$,
\item[\bf 7)] If $t\leq N$ return to {\bf 3)}.
\end{itemize}
}
\end{minipage}
}

A generic choice for $q(\btheta|\btheta^{(t-1)})$ is the random walk proposal: 
$q(\btheta|\btheta^{(t-1)})=g(\btheta-\btheta^{(t-1)})$ with a symmetric function $g$,
which provides a simplified acceptance probability. Indeed, in that case, step {\bf 5)} of the previous algorithm
is replaced with: if 
$$
u \leq \frac{\pi(\btheta')f(\by|\btheta')}{\pi(\btheta^{(t-1)})f(\by|\btheta^{(t-1)})}\,,
$$
set $\btheta^{(t)}=\btheta'$ else $\btheta^{(t)}=\btheta^{(t-1)}$. \\
This ensures that values $\btheta'$ that are more likely 
than the current $\btheta^{(t-1)}$ are always accepted while values that are
less likely are sometimes accepted.

\begin{figure*}
\begin{center}
 \includegraphics[width=0.4\textwidth]{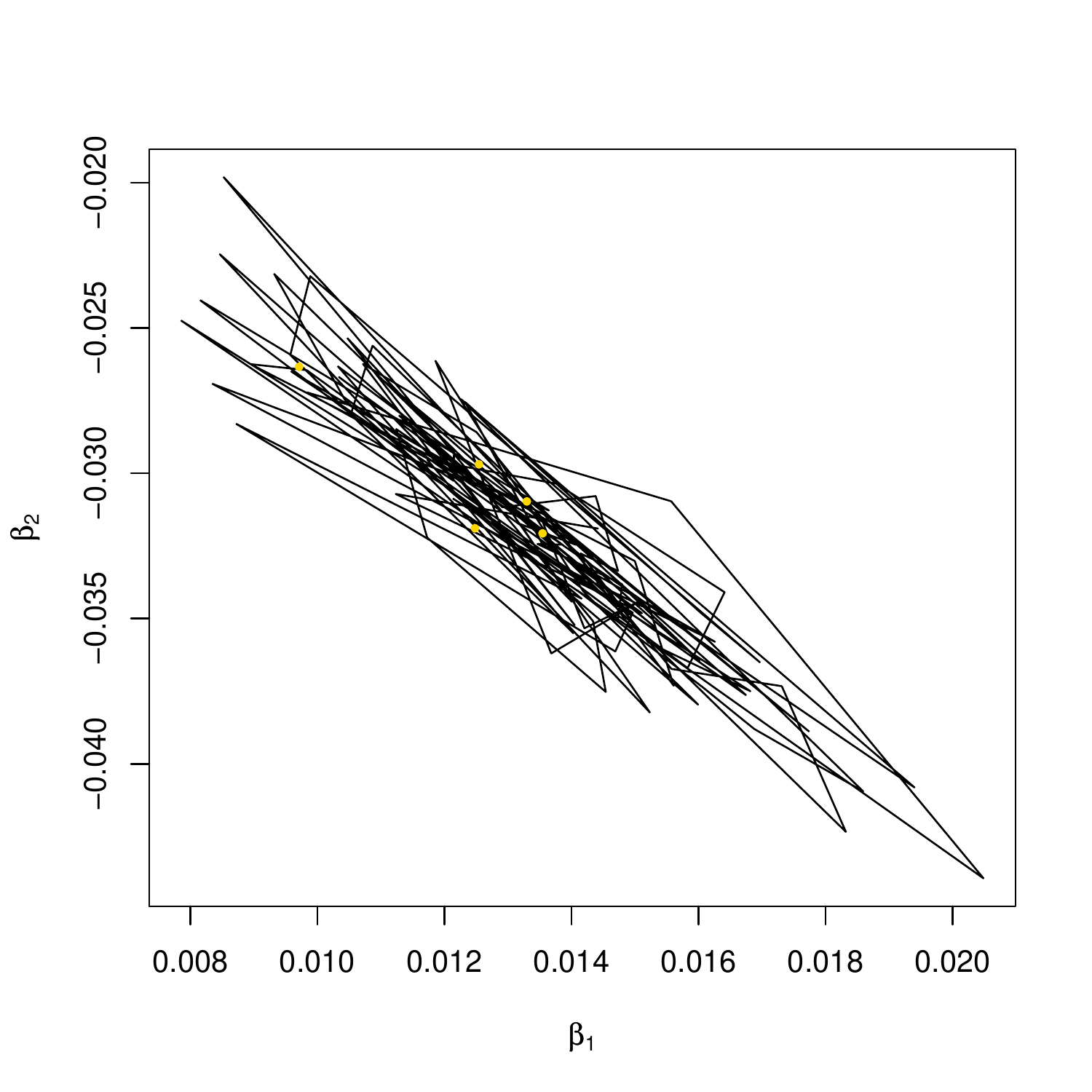}
 \includegraphics[width=0.4\textwidth]{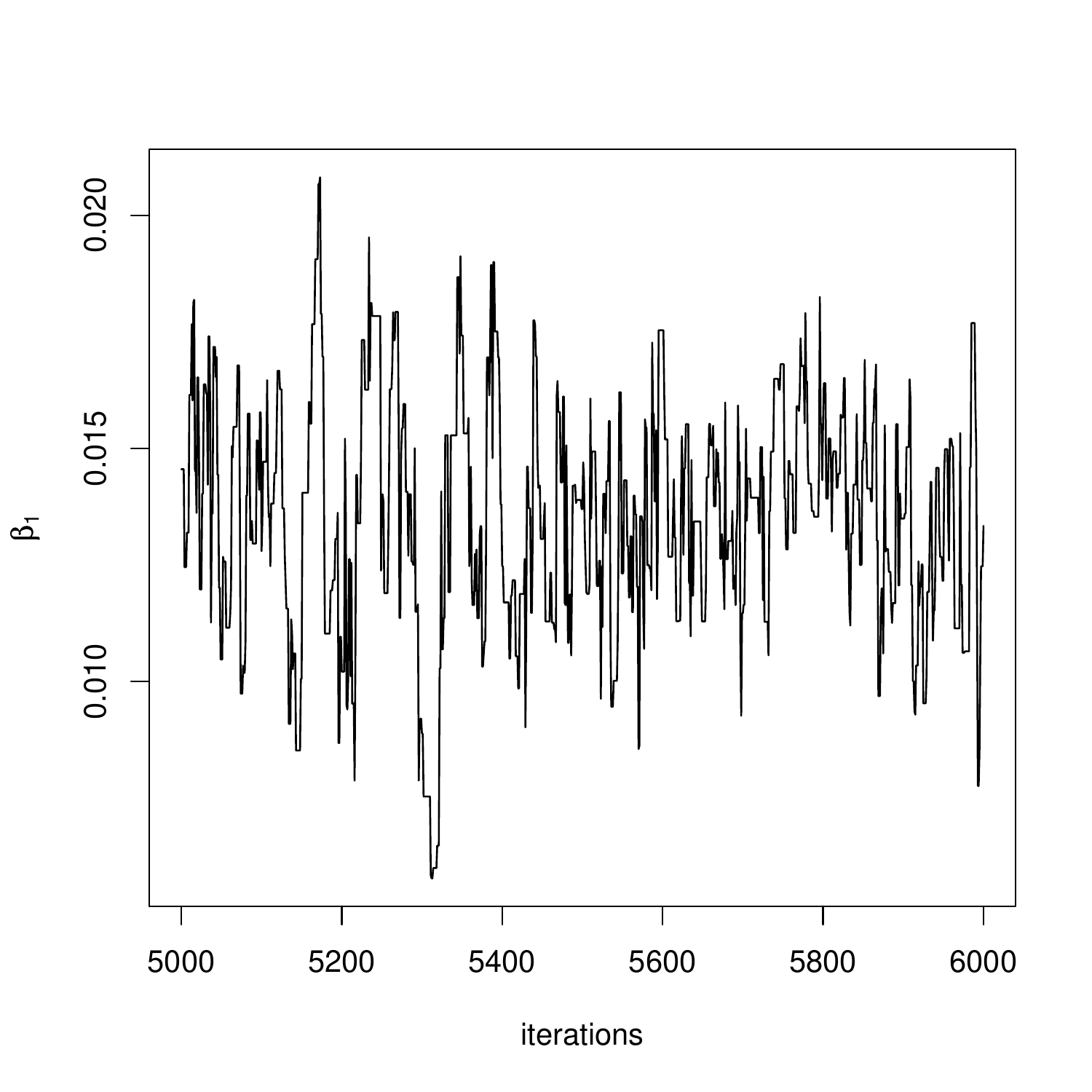}
\end{center}
\caption{\label{fig:mhpro}
Random-walk Metropolis--Hastings algorithm applied to the Pima Indian diabetes dataset. The left graph describes the path of
the subchain $(\bbeta^{(100t)})$.
The right graph shows the path of the first component chain $\beta_1^{(t)}$ for
$5000\le t\le 6000$. 
}
\end{figure*}

\begin{example}[Continuation of Example \ref{ex:glum}]\label{ex:glumig}\begin{rm}
If we consider the Pima Indian diabetes dataset with only its first two covariates, the parameter
$\bbeta$ is of dimension $2$ and the random walk proposal can be easily implemented. We use for $g$
a normal distribution with covariance matrix the asymptotic covariance matrix $\hat\Sigma$ of the MLE
and the proposed value $\bbeta'$ is then simulated at iteration $t$ as
$$
\bbeta'\sim\mathcal{N}_2(\beta^{(t-1)},\hat\Sigma)\,.
$$
The MLE may also be used as starting value for the chain.
Figure \ref{fig:mhpro} illustrates the behaviour of the Metropolis--Hastings algorithm for this dataset, the lhs graph
describing the path of the subchain $(\bbeta^{(100t)})$ and the rhs detailing the first component $\beta_1^{(t)}$ for 
$5000\le t\le 6000$. Although this is not clearly visible on the rhs graph, the acceptance rate of the algorithm is close to
$50\%$, which means that half of the proposed $\bbeta$'s are rejected.\footnote{This rate happens to be almost optimal for
small dimensions \citep{gelman:gilks:roberts:1996}.} Using a covariance matrix that is five times larger leads to an
acceptance rate of $25\%$, while the larger $10\hat\Sigma$ produces an acceptance rate of $15\%$.
\findeX\end{example}

Finding the proper scale is not always as straightforward as in Example \ref{ex:glumib} and asymptotic normal approximations
to the posterior distribution may be very inefficient. While the Metropolis--Hastings algorithm recovers better from facing
large-dimensional problems than standard importance sampling techniques, this still is a strong limitation to its use in
large-dimensional setups.

\subsection{Gibbs sampling}
In contrast, the alternative Gibbs sampler is an attractive algorithm for large-dimensional problems
because it naturally fits the hierarchical structures often present in 
Bayesian models and more generally in graphical and latent variable models.
The fundamental strength of the Gibbs sampler is its ability to
break a joint target distribution like $\pi(\theta_1,\ldots,\theta_p|\by)$ in 
the corresponding conditional distributions $\pi_i(\theta_i|\by,\btheta_{-i})$ 
$(i=1,\ldots,n)$ and to simulate successively from these low-dimensional targets:

\bigskip
\noindent \fbox{
\begin{minipage}{0.46\textwidth}
{\bf $p$-component systematic scan Gibbs sampler}\\
{\sf
For a computing effort $N$
\begin{itemize}
\item[\bf 1)] Choose $\btheta^{(0)}$,
\item[\bf 2)] Set $t=1$,
\item[\bf 3)] Generate $\theta_1^{(t)}$ from $\pi_1\left(\theta_1|\by,\btheta_{-1}^{(t-1)}\right)$,
\item[\bf 4)] Generate $\theta_2^{(t)}$ from $\pi_2\left(\theta_2|\by,\theta_1^{(t)},\btheta_{-(1:2)}^{(t-1)}\right)$,
\item[\bf 5)] \ldots
\item[\bf 6)] Generate $\theta_p^{(t)}\sim \pi_p\left(\theta_p|\by,\btheta_{(1:(p-1))}^{(t)}\right)$, 
\item[\bf 7)] Set $t=t+1$,
\item[\bf 8)] If $t\leq N$ return to {\bf 3)}.
\end{itemize}
}
\end{minipage}
}

\bigskip
While this algorithm seems restricted to mostly hierarchical multidimensional models, the
special case of the {\em slice sampler} (Robert and Casella, 2004, Chapter 8) shows that
the Gibbs sampler applies in a wide variety of models. 

\begin{figure*}
\begin{center}
 \includegraphics[width=0.4\textwidth]{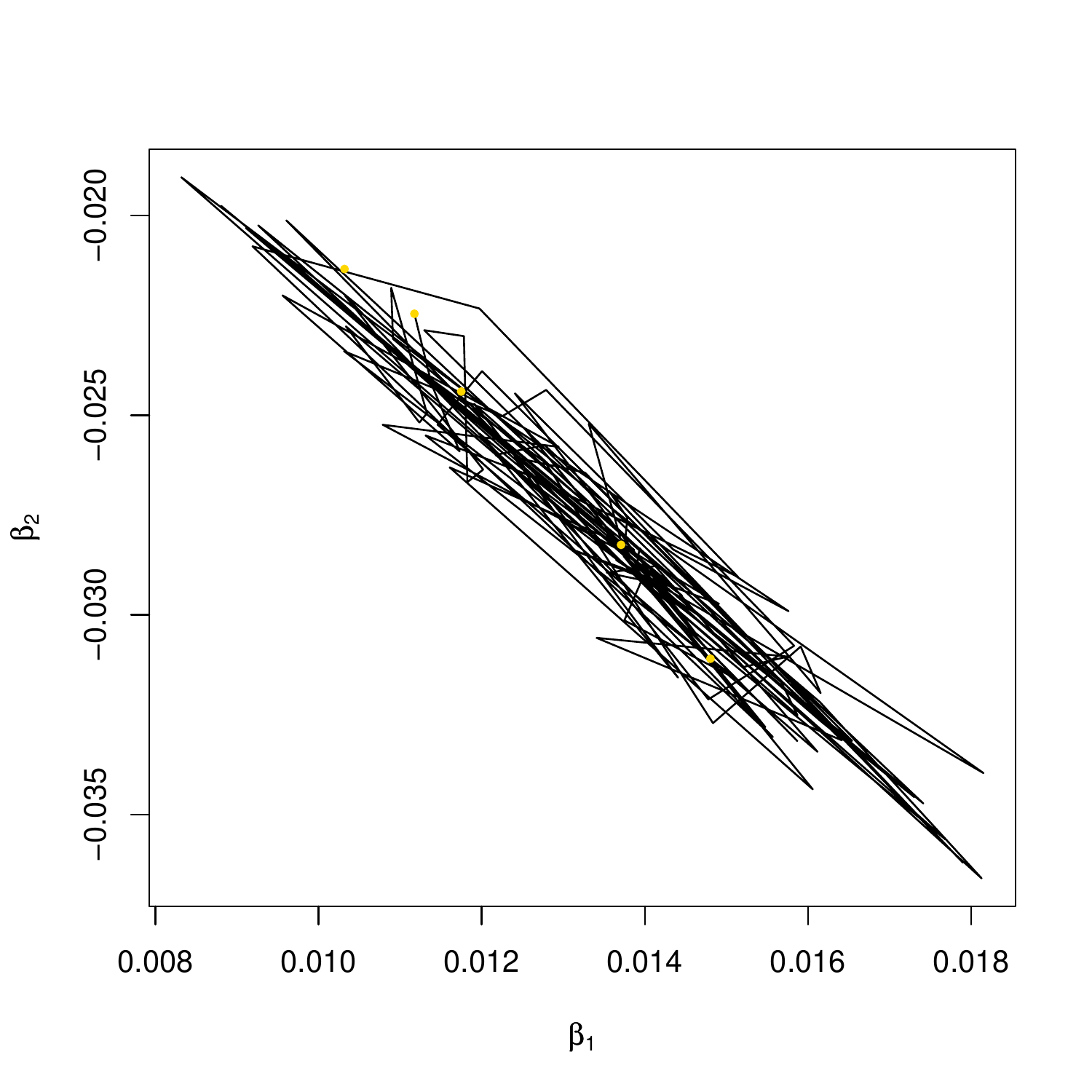}
 \includegraphics[width=0.4\textwidth]{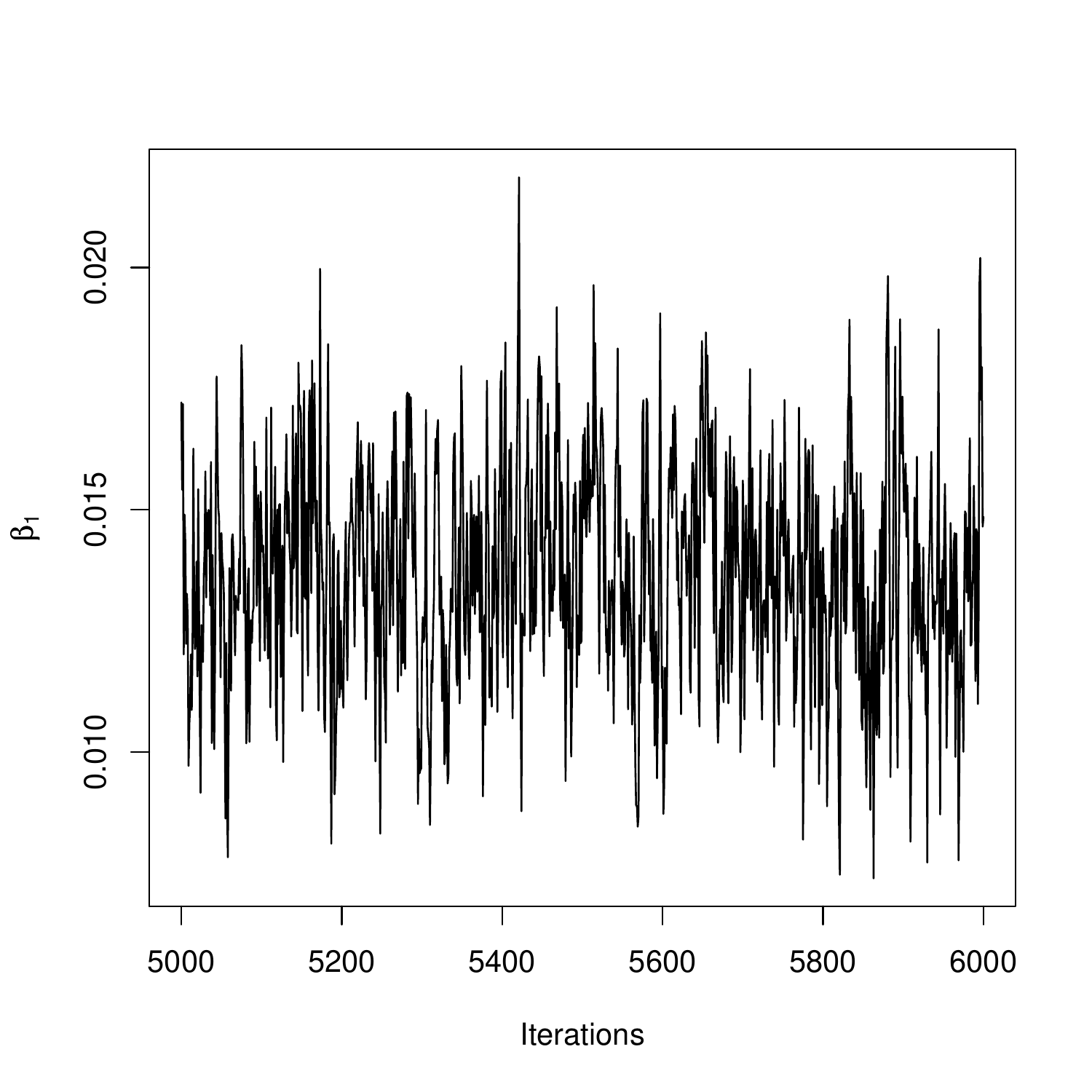}
\end{center}
\caption{\label{fig:gibpro}
Gibbs sampling algorithm applied to the Pima Indian diabetes dataset. The left graph describes the path of
the subchain $(\bbeta^{(100t)})$.
The right graph shows the path of the first component chain $\bbeta_1^{(t)}$ for $5000\le t\le 6000$.
}
\end{figure*}

\begin{example}\label{probbs}\begin{rm} (Continuation of Example \ref{probitatent})
As noted in Example \ref{probitatent}, the probit model allows for a latent variable
representation based on the artificial normal variable $z_t$ connected with the observed
variable $y_t$. This representation opens the door to a Gibbs sampler \citep{albert:chib:1993b}
aimed at the joint posterior distribution of $(\bbeta,\bz)$ given $\by$. Indeed, the conditional 
distribution of the latent variable $z_t$ given $\bbeta$ and $y_t$,
\begin{equation}
z_t|y_t,\bbeta\sim\left\{\begin{array}{ll}
\mathcal{N}_+\left(\bx_t^\text{T}\bbeta,1,0\right) & \mbox{if}\quad y_t=1\,, \\
\mathcal{N}_-\left(\bx_t^\text{T}\bbeta,1,0\right) & \mbox{if}\quad y_t=0\,,
\end{array}\right.
\label{gibbs1}
\end{equation}
is clearly available,\footnote{Here, 
$\mathcal{N}_+\left(\bx_t^\text{T}\bbeta,\allowbreak 1,\allowbreak 0\right)$ denotes the normal
distribution with mean $\bx_t^\text{T}\bbeta$ and variance $1$ that is left-truncated at $0$, while
$\mathcal{N}_-\left(x_t^\text{T}\bbeta,1,0\right)$ denotes the symmetrical normal distribution that 
is right-truncated at $0$.}. The corresponding full conditional on the parameters 
is given by the standard normal distribution (which does not depend on $\by$)
\begin{align}
\bbeta|\bz\sim\mathcal{N}&\left(\frac{n}{n+1}(\bX^\text{T}\bX)^{-1}\bX^\text{T}\bz,\right.\nonumber\\
&\left.\frac{n}{n+1}(\bX^\text{T}\bX)^{-1}\right)\,.
\label{gibbs2}
\end{align}
Therefore, given the current value of $\bbeta$, one cycle of the Gibbs algorithm
produces a new value for $\bz$  as simulated from the
conditional distribution (\ref{gibbs1}), which,
when substituted into (\ref{gibbs2}), produces a new value for $\bbeta$.
Although it does not impact the long-term properties of the sampler,
the starting value of $\bbeta$ may once again be taken as the maximum likelihood estimate
to avoid (useless) burning steps in the Gibbs sampler.

The implementation of this Gibbs sampler is straightforward. There is no parameter to calibrate
(as opposed to the scale in the random-walk Metropolis--Hastings scenario). When comparing Figure \ref{fig:mhpro}
and \ref{fig:gibpro}, the raw plot of the sequence
$(\bbeta_1^{(t)})$ shows that the mixing behaviour of the Gibbs sampling chain is superior to the one for the
Metropolis--Hastings chain.
\findeX\end{example}

\subsection{Hybrid solutions}
Mixing both Metropolis--Hastings and Gibbs algorithms often result in better performances
like faster convergence of the resulting Markov chain, the former algorithm being often 
used for global exploration of the target and the later for local improvement.

A classic hybrid algorithm replaces a non-available Gibbs update by
a Metropolis--Hastings step. Another hybrid solution alternates Gibbs and
Metropolis--Hastings proposals. The corresponding algorithms are valid:
they produce ergodic Markov chains with the posterior target as stationary distribution.

\begin{figure}
\includegraphics[width=4.5cm,angle=270]{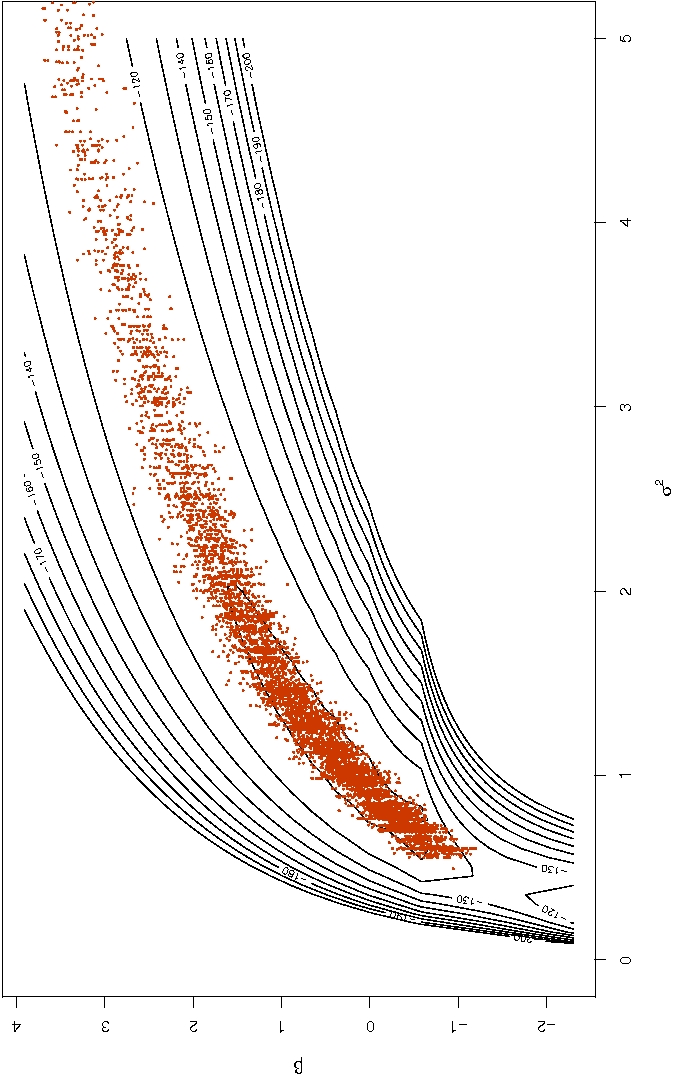} 
\caption{\label{fig:newpro1}
Contour plot of the log-posterior distribution for a probit sample of $1,000$
observations, along with $1,000$ points of an MCMC sample {\em (Source:
\citealp{Robert:Casella:2004})}.}
\end{figure}

\begin{example}[Continuation of Example \ref{ex:glumbit}]\label{ex:gimli}
\begin{rm}
For $p=1$, the probit model can be over-parameterised as
$$
\P(Y_i=1|x_i)=1-\P(Y_i=0|x_i) = \Phi(x_i\beta/\sigma)\,,
$$
while only depending on $\beta/\sigma$. Using a proper prior like
\begin{align*}
\pi(\beta,\sigma^2|\bx)&=\pi(\beta|\bx)\pi(\sigma^2|\bx) \\
&\propto \sigma^{-4}\,\exp\{-1/\sigma^2\}\,\exp\{-\beta^2/50)\,,
\end{align*}
the corresponding Gibbs sampler simulates $\beta$ and $\sigma^2$ alternatively, from
$$
\pi(\beta|\bx,\by,\sigma) \propto \prod_{i=1}^n \Phi(x_i\beta/\sigma)^{y_i}\Phi(-x_i\beta/\sigma)^{1-y_i} \pi(\beta|\bx)
$$
and
$$
\pi(\sigma^2|\bx,\by,\beta) \propto \prod_{i=1}^n \Phi(x_i\beta/\sigma)^{y_i}\Phi(-x_i\beta/\sigma)^{1-y_i} \pi(\sigma^2|\bx)
$$
respectively.  Since both of these conditional distributions are non-standard, we
replace the direct simulation by one-dimensional Metropolis--Hastings steps,\footnote{In this Metropolis-within-Gibbs
strategy, note that a {\em single} step of a Metropolis--Hastings move is sufficient to validate the algorithm, since
stationarity, not convergence, is the issue.} using
normal $\mathcal{N}(\beta^{(t)},1)$ and log-normal $\mathcal{L}\mathcal{N}(\log\sigma^{(t)},.04)$
random walk proposals, respectively. (The scales were found by trial-and-error.) For a simulated dataset of $1,000$ points,
the contour plot of the log-posterior distribution is given in Figure \ref{fig:newpro1},
along with the last $1,000$ points of a corresponding MCMC sample after $100,000$ iterations.
This graph shows a very satisfactory repartition of the simulated parameters over the likelihood
surface, with higher concentrations near the largest posterior regions. 
\findeX\end{example}

Let us note as a conclusion to this short section that an alternative meaning for hybrid solutions is the
simultaneous use of different Markov kernels \citep{tierney:1994}. A mixture of MCMC kernels does remain an
MCMC kernel with the same stationary distribution and its performances are at least as good as the best component
in the mixture. There is therefore very little to say against advocating this extension.

\subsection{Scaling and adaptivity}\label{scala}
A difficulty with Metropolis--Hastings algorithms, including random walk versions,
is the calibration of the proposal distribution: this proposal must be sufficiently
related to the target distribution so that, in a reasonable number of steps, the
whole support of this distribution can be visited. If the scale of the
random walk proposal is too small, this will not happen as the algorithm stays ``too
local" and, if for instance there are several modes on the target, the algorithm
may remain trapped within one modal region because it cannot reach other
modal regions with jumps of too small a magnitude. 

\begin{figure}
\begin{center}
\includegraphics[height=5cm,width=7cm]{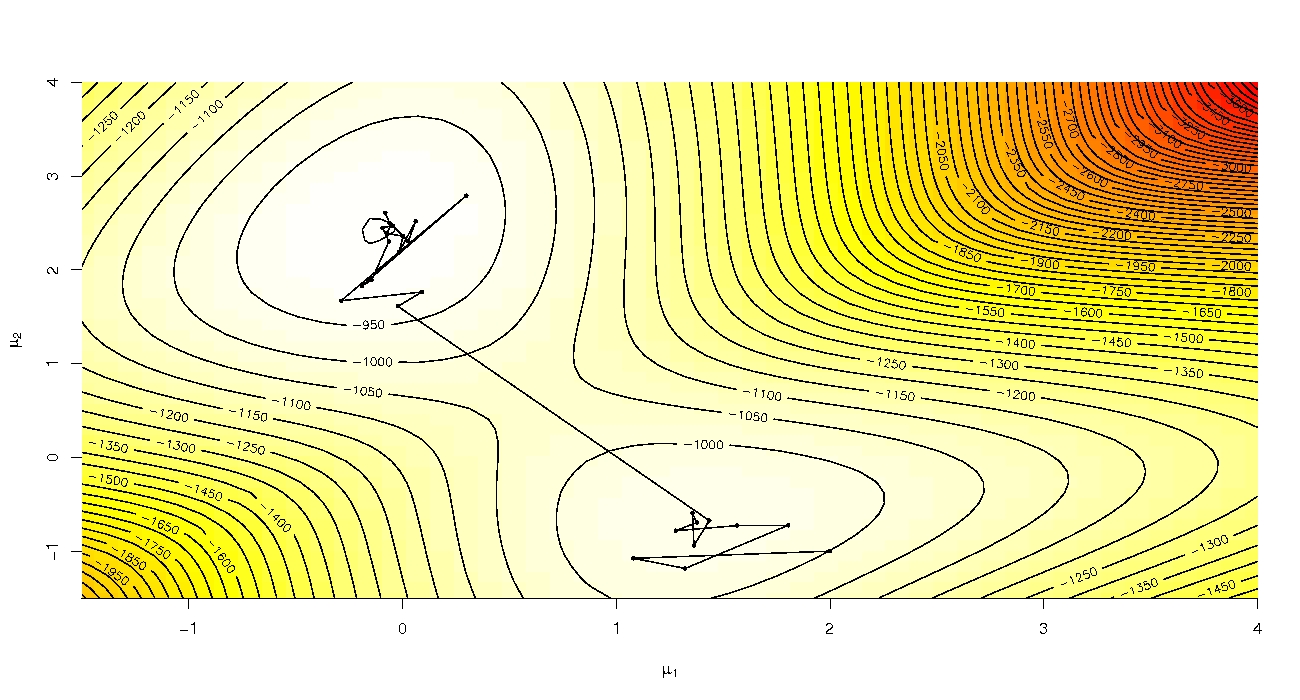}
\includegraphics[height=5cm,width=7cm]{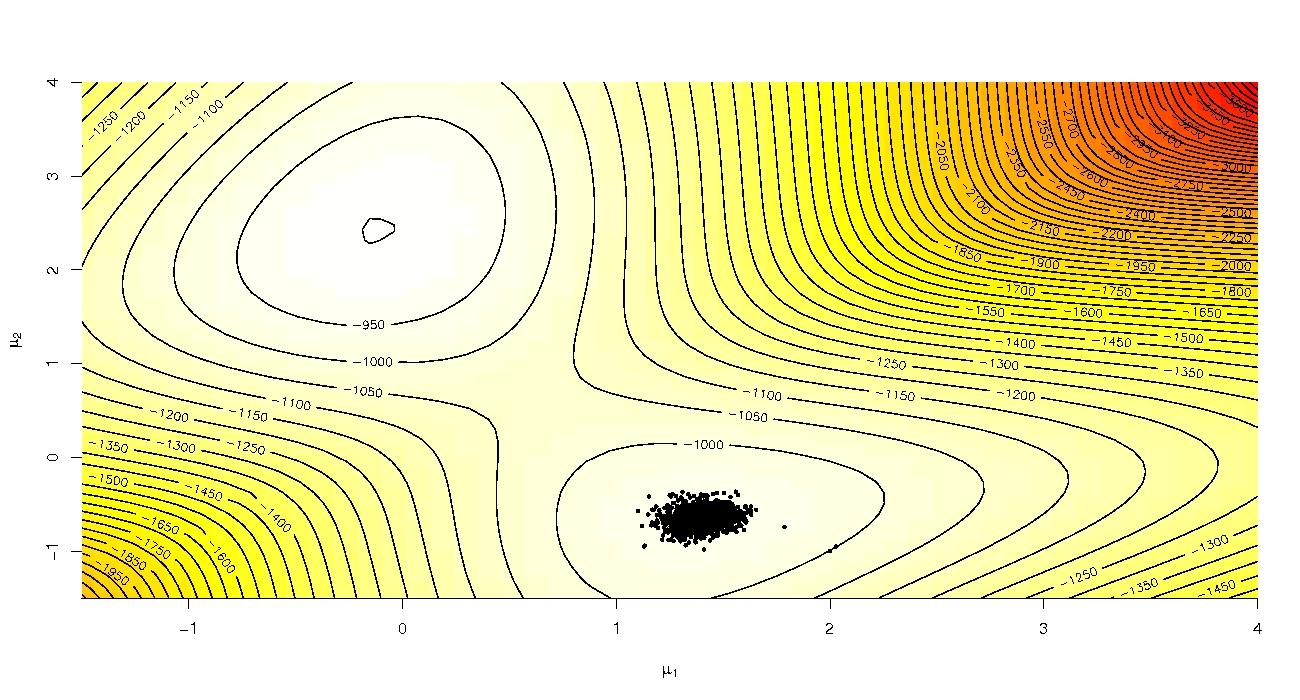}
\end{center}
\caption{\label{fig:lo&hi} Evolution of a random walk Metropolis--Hastings chain on
a mixture log-posterior surface for $n=500$ observations 
and {\em (top)} $\tau=1$ and $1,000$ iterations;
 {\em (bottom)} $\tau=.3$ and $10,000$ iterations.}
\end{figure}

\begin{example}\label{mixomu}\begin{rm} For a sample $y_1,\ldots,y_n$ from the mixture distribution
$$
p\,\mathcal{N}(\mu_1,\sigma^2) + (1-p)\,\mathcal{N}(\mu_2,\sigma^2) \,
$$
where both $p$ and $\sigma^2$ are known,
the posterior distribution associated with the prior $\mathcal{N}(0,10\sigma^2)$
on both $\mu_1$ and $\mu_2$ is multimodal, with a major mode close to the true value
of $\mu_1$ and $\mu_2$ (when $n$ is large enough) and a secondary and spurious mode
(that stems from the nonindentifiable case $p=0.5$).
When running a random walk Metropolis--Hastings algorithm on this model, with a
normal proposal $\mathscr{N}_2((\mu_1^{(t)},\mu_2^{(t)}),\tau \mathbf{I}_2)$,
a small scale $\tau$ prevents the Markov chain from visiting the major mode.
Figure \ref{fig:lo&hi} compares two choices of $\tau$ for the same dataset: for
$\tau=1$, the spurious mode can be escaped but for $\tau=.3$ the chain remains trapped
in that starting mode.
\findeX\end{example}

The larger the dimension $p$ is, the harder the determination of the scale is, because
\begin{enumerate}\renewcommand{\theenumi}{\alph{enumi}}
\item the curse of dimensionality implies that there is an increasingly important part of
the space with zero probability under the target;
\item the knowledge and intuition about the modal regions get weaker (for complex
distributions, it is impossible to identify none but a few of the modes);
\item the proper scaling of a random walk proposal
involves a symmetric $(p,p)$ matrix. Even when diagonal, this matrix gets harder to scale as the
dimension increases (unless one resorts to a Gibbs like implementation,
where each direction is scaled separately).
\end{enumerate} 

\begin{figure}
\includegraphics[height=5cm,width=8cm]{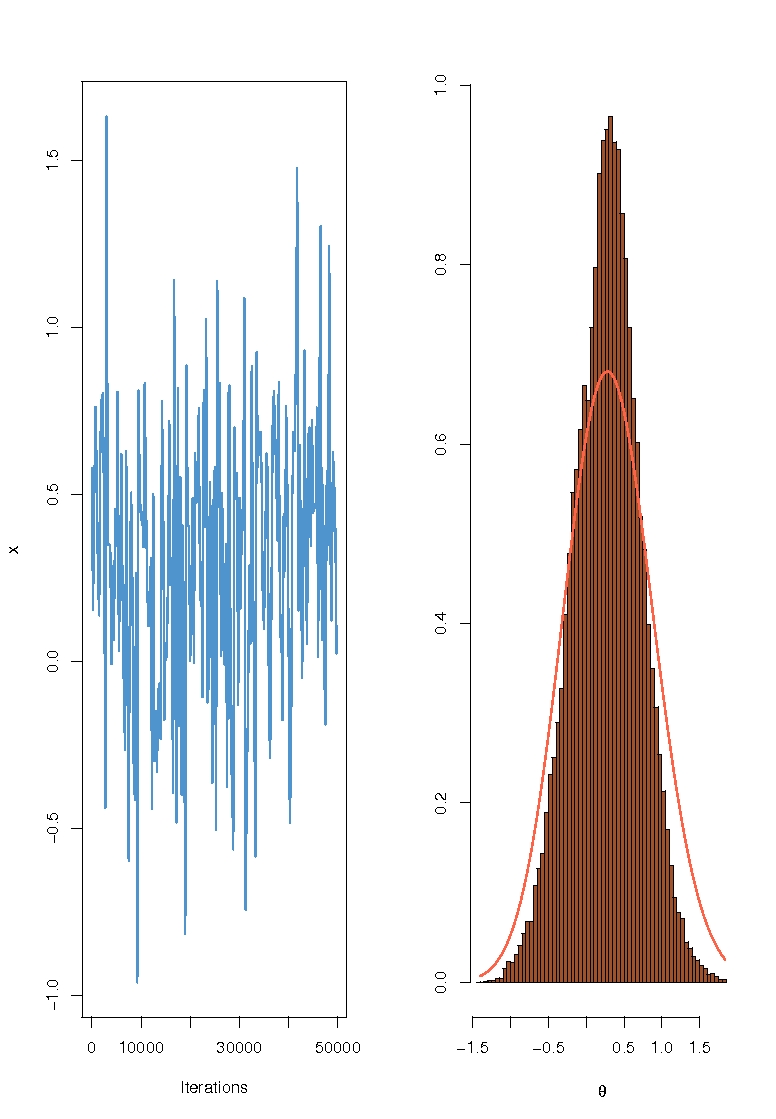} 
\caption{\label{fig:nodapt} Sample {\em (left)} produced by $50,000$ iterations of 
a nonparametric adaptive MCMC scheme and comparison {\em (right)} of its distribution 
with the target distribution {\em (solid curve)}. {\em (Source: \citealp{Robert:2003})}}
\end{figure}

In addition to these difficulties, learning about the specificities of the target
distribution while running an MCMC algorithm and tuning the proposal accordingly,
i.e.~constructing an adaptive MCMC procedure, is difficult because this 
cancels the Markov property of the original method and thus jeopardizes convergence.
For instance, Figure \ref{fig:nodapt} shows the discrepancy between an histogram of
a simulated Markov chain and the theoretical limit (solid curve) when the proposal distribution
at time $T$ is a kernel approximation based on the first $T-1$ simulations of the ``chain". 
Similarly, using an on-line scaling of the algorithm against the empirical acceptance 
rate in order to reach a golden number like $0.234$ \citep[see][Note 7.8.4]{Robert:Casella:2004}
is inherently flawed in that the attraction of a modal region may give a false sense 
of convergence and may thus lead to a choice of too small a scale, simply because other modes 
will fail to be visited during the scaling experiment.

However, there are algorithms that preserve ergodicity (convergence to the
target) while implementing adaptivity. See, e.g.,
\cite{Gilks:Roberts:Sahu:1998} who use regeneration to create block independence and
preserve Markovianity on the paths rather than on the values, \cite{Haario:Sacksman:Tamminen:1999,
Haario:Sacksman:Tamminen:2001} who derive a proper adaptation scheme 
by using a ridge-like correction to the empirical variance in very large dimensions
for satellite imaging data, and \cite{Andrieu:Moulines:Priouret:2005}
who propose a general framework of valid adaptivity based on stochastic optimisation and the Robbin-Monro
algorithm.

More recently, \cite{roberts:rosenthal:2007} consider basic ergodicity properties of adaptive Markov chain Monte Carlo algorithms
under minimal assumptions, using coupling constructions. They prove convergence in distribution and a weak law of large numbers.
Moreover, in \cite{roberts:rosenthal:2006}, they investigate the use of adaptive MCMC algorithms to automatically tune the Markov
chain parameters during a run. Examples include the adaptive Metropolis multivariate algorithm of \cite{Haario:Sacksman:Tamminen:2001},
Metropolis-within-Gibbs algorithms for nonconjugate hierarchical models, regionally adjusted Metropolis algorithms, and logarithmic scalings.
\cite{roberts:rosenthal:2006} present some computer simulation results that indicate that the algorithms 
perform very well compared to non-adaptive algorithms, even in high dimensions.

\section{Approximate Bayesian computation techniques}
There exist situations where the likelihood function $f(\by|\btheta)$ is overly expensive or even impossible to calculate,
but where simulations from the density $f(\by|\btheta)$ are reasonably produced. A generic class of such situations
is made by latent variable models where the analytic integration of the latent variables is impossible, while handling
the latent variables as additional parameters in a joint distribution causes any MCMC to face convergence problems. Another
illustration is given by inverse problems where computing the function $f(\by|\btheta)$ for a given pair $(\by,\btheta)$
involves solving a numerical equation. In such cases,
it is almost impossible to use the computational tools presented in the previous section
to sample from the posterior distribution $\pi(\btheta|\by)$. Approximate Bayesian computation (ABC) 
is an alternative to such techniques that only requires being able to sample from the likelihood $f(\cdot|\btheta)$.
It was first proposed for population genetic models  \citep{beaumont:zhang:balding:2002} but applies in much wider
generality.

\bigskip
\noindent \fbox{
\begin{minipage}{0.46\textwidth}
{\bf Likelihood free rejection sampling}\\
{\sf
For a computing effort $N$
\begin{itemize}
\item[\bf 1)] Set $i=1$,
\item[\bf 2)] Generate $\btheta'$ from the prior distribution $\pi(\cdot)$,
\item[\bf 3)] Generate $\bz$ from the likelihood $f(\cdot|\btheta')$,
\item[\bf 4)] If $\rho(\eta(\bz),\eta(\by))\leq \epsilon$, set $\btheta_i=\btheta'$ and $i=i+1$,
\item[\bf 5)] If $i\leq N$, return to {\bf 2)}.
\end{itemize}
}
\end{minipage}
}

\bigskip
This likelihood free algorithm samples from the marginal in $\bz$ of the following joint distribution:
$$
\pi_\epsilon(\btheta,\bz|\by) \propto
{\pi(\btheta)f(\bz|\btheta)\mathbb{I}_{P^{\epsilon,\by}}(\bz)}
$$
with the tuning parameters as
\begin{itemize}
\item $\rho(\cdot,\cdot)$ a distance,
\item $\eta(\cdot)$ a summary statistic,
\item $\epsilon>0$ a tolerance level,
\item $P^{\epsilon,\by}=\{\bz|\rho(\eta(\bz),\eta(\by))<\epsilon\}$.
\end{itemize}
The idea behind ABC \citep{beaumont:zhang:balding:2002} is that the summary statistics coupled with a small tolerance should
provide a good approximation of the posterior distribution:
$$
\pi_\epsilon(\btheta|\by)=\int \pi_\epsilon(\btheta,\bz|\by)\text{d}\bz\approx \pi(\btheta|\by)\,.
$$

It has been shown in \cite{marjoram:etal:2003} that it is possible to construct a Metropolis--Hastings algorithm
that samples from $\pi_\epsilon(\btheta,\bz|\by)$, and the marginally from $\pi_\epsilon(\btheta|\by)$;
this algorithm only requires the ability to sample from $f(\cdot|\btheta)$. This is the likelihood free MCMC sampler:

\bigskip
\noindent \fbox{
\begin{minipage}{0.46\textwidth}
{\bf Likelihood free MCMC sampler}\\
{\sf
For a computing effort $N$
\begin{itemize}
\item[\bf 1)] Use the likelihood free rejection sampling to get a realization $\btheta^{(0)}$ from the ABC target distribution $\pi_\epsilon(\btheta|\by)$,
\item[\bf 2)] Set $t=1$,
\item[\bf 3)] Generate $\btheta'$ from the Markov kernel $q\left(\cdot|\btheta^{(t-1)}\right)$,
\item[\bf 4)] Generate $\bz$ from the likelihood $f(\cdot|\btheta')$,
\item[\bf 5)] Generate $u$ from $\mathcal{U}_{[0,1]}$,
\item[\bf 6)] If $u \leq 
\dfrac{\pi(\btheta')q(\btheta^{(t-1)}|\btheta')}{\pi(\btheta^{(t-1)})q(\btheta'|\btheta^{(t-1)})}\mathbb{I}_{P^{\epsilon,\by}}(\bz)$, \\
set $\btheta^{(t)}=\btheta'$ else $\btheta^{(t)}=\btheta^{(t-1)}$,
\item[\bf 7)] Set $t=t+1$,
\item[\bf 8)] If $t\leq N$ return to {\bf 3)}.
\end{itemize}
}
\end{minipage}
}

\bigskip
Rejection sampling and MCMC methods can perform poorly if the tolerance level $\epsilon$ is small.
Consequently various sequential Monte Carlo algorithms have been constructed as an alternative to 
both methods. For instance, \cite{beaumont:cornuet:marin:robert:2009} proposed an ABC version of the
Population Monte Carlo algorithm presented above. The key idea is to decompose the difficult issue
of sampling from $\pi_\epsilon(\btheta,\bz|\by)$ into a series of simpler subproblems. The algorithm
begins at time 0 sampling from $\pi_{\epsilon_0}(\btheta,\bz|\by)$ with a large value $\epsilon_0$, then
simulating from an increasing difficult sequence of target distribution $\pi_{\epsilon_t}(\btheta,\bz|\by)$,
that is when $\epsilon_t<\epsilon_{t-1}$.

\begin{figure}
\centerline{\includegraphics[width=7cm]{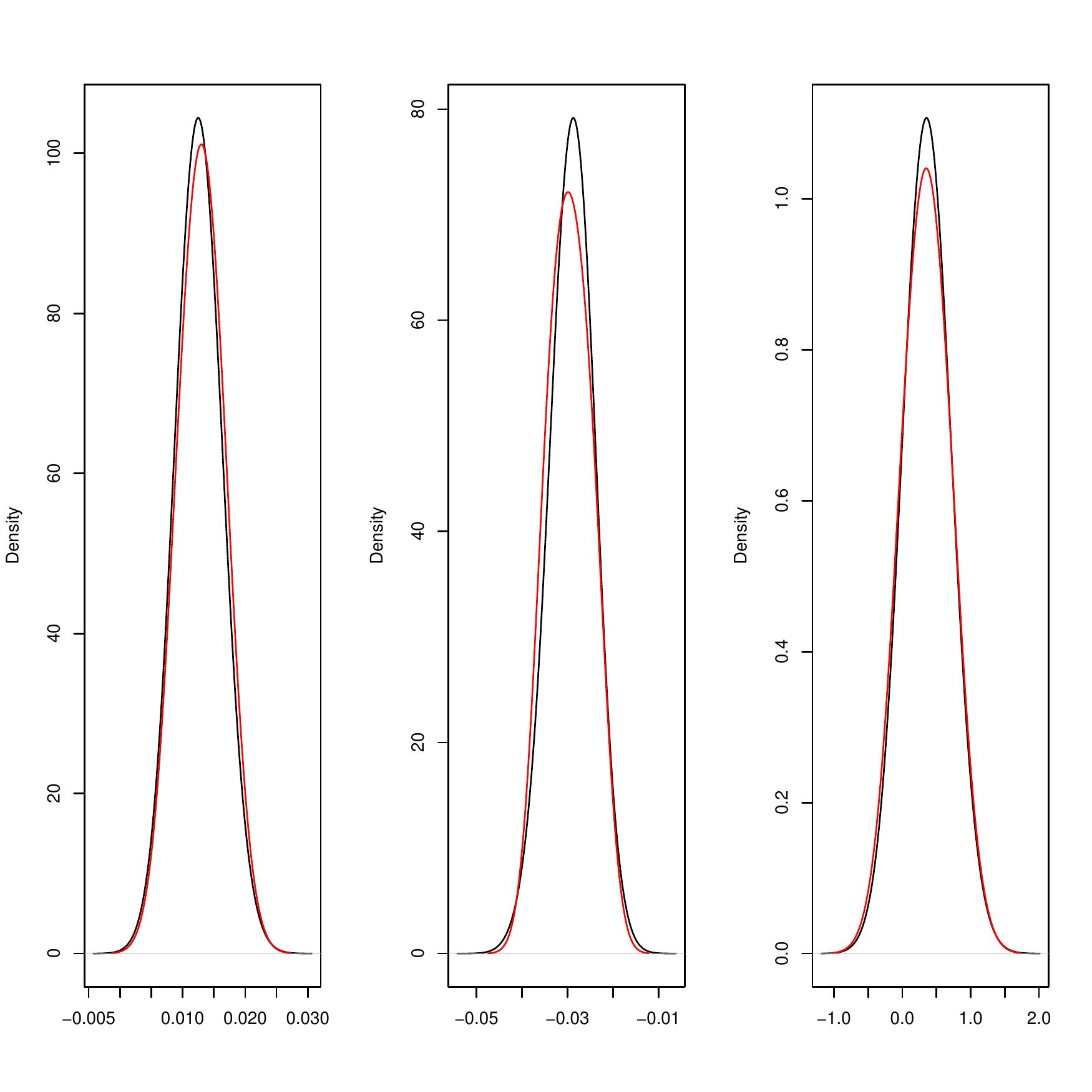}}
\caption{\label{fig:abc} 
Comparison between density estimates of the marginal posterior distributions of
$\beta_1$ (left), $\beta_2$ (center) and $\beta_3$ (right) obtained by ABC (in red) and MCMC samples (in black)
in the setup of the Pima Indian diabetes study.}
\end{figure}

\begin{example}[Continuation of Example \ref{ex:glumib}]\label{ex:glumabc}\begin{rm}
Figure \ref{fig:abc} provides an illustration of the above algorithm when applied to the probit
model with the three covariates described in Example \ref{ex:glum}. In this artificial case, the
ABC outcome can be compared with the MCMC ``exact" simulation described above and the result is striking in that
the ABC approximation is confounded with the exact posterior densities. The tuning of the ABC algorithm is
to use $10^6$ simulations over $10$ iterations, with bounds $\epsilon_t$ set as the $1\%$ quantile of the simulated
$\rho(\eta(\bz),\eta(\by))$, $\rho$ chosen as the Euclidean distance, and $\eta(\bz)$ as the predictive distribution based
on the current parameter $\beta$, made of the $\Phi(x_i^\Tee\beta)$'s, 
while $\eta(\by)$ is the predictive distribution based on the MLE $\hat\beta(\by)$
made of the $\Phi(x_i^\Tee\hat\beta(\by))$'s.
In this special case we are therefore avoiding the simulation of the observations themselves as predictive functions are
available. This choice reduces the variability in the divergence between $\eta(\bz)$ and $\eta(\by)$, and explains for
the very good fit between the densities.
\findeX\end{example}

\section{Final remarks}
This tutorial is necessarily incomplete and biased: the insistance on model choice and on variable
dimension models is also a reflection of the author's own interests. Others would have 
rather chosen to stress the relevance of these simulation methods for optimal design 
\citealp{Mueller:1999, Mueller:Parmigiani:Robert:Rousseau:2004} in
conjonction with simulated annealing \citep[e.g.][]{Andrieu:Doucet:2000,Doucet:Godsill:Robert:2002},
for non-parametric regression \citep{Denison:Holmes:Mallick:Smith:2002} or for the analysis of
continuous time stochastic processes \citep{Roberts:Papaspiliopoulos:Dellaportas:2001,
Beskos:Papaspiliopoulos:Roberts:Fearnhead:2006}. That such a wealth of choices is available 
indicates that the field still undergoes a formidable expansion that should benefit a wide
range of areas and disciplines and, conversely, that the continued attraction of new areas
within the orbit of Bayesian computational methods backfeeds their creativity by introducing
new challenges and new paradigms.

\section*{Acknowledgement}
This work had been partly supported by the Agence Nationale de la Recherche (ANR, 212,
rue de Bercy 75012 Paris) through the 2009-2012 projects {\sf Big'MC} and {EMILE}.

\bibliographystyle{ims}

\end{document}